\documentclass[11pt,double]{article}
\usepackage[latin1]{inputenc}
\usepackage{amscd}
\usepackage{amsmath}
\usepackage{amssymb}
\usepackage{amsfonts}
\usepackage{latexsym}
\usepackage{amsthm}
\usepackage{mathrsfs}
\usepackage{booktabs}
\usepackage{caption}
\usepackage{mathtools} 
\usepackage[nohead]{geometry}
\usepackage{epsfig}
\usepackage{algorithm}		
\usepackage{authblk}
\usepackage{cite}
\usepackage{helvet}         
\usepackage{courier}        
\usepackage{url}
\usepackage{algpseudocode}
\usepackage{algpascal}
\usepackage{multirow}
\usepackage{setspace}
\numberwithin{equation}{section}

\newtheorem{theorem}{Theorem}[section]
\newtheorem{definition}{Definition}[section]
\newtheorem{corollary}{Corollary}[section]
\newtheorem{proposition}{Proposition}[section]

\def\bdelta{\mbox{\boldmath $\delta$}}

\def\bmu{\mbox{\boldmath $\mu$}}

\def\bOmega{\mbox{\boldmath $\Omega$}}

\def\bpsi{\mbox{\boldmath $\psi$}}
\def\bphi{\mbox{\boldmath $\phi$}}

\def\bLambda{\mbox{\boldmath $\Lambda$}}

\def\bSigma{\mathbf{\Sigma}}

\def\bb{\mathbf{b}}

\def\bff{\mbox{\boldmath $f$}}

\def\by{\mathbf{y}} 
\def\bY{\mathbf{Y}}
\def\0{\mbox{\bf{0}}}

\def\bz{\mathbf{z}}
\def\bZ{\mathbf{Z}}
\def\bC{\mathbf{C}}

\def\bA{\mathbf{A}}

\def\bQ{\mathbf{Q}}

\def\bB{\mathbf{B}}

\def\sd{\mathsf{d}}

\def\sp{\mathsf{p}} 

\def\sw{\mathsf{w}}

\def\sy{\mathsf{y}} 
\def\sY{\mathsf{Y}}

\def\sn{\mathsf{n}}

\def\sz{\mathsf{z}}
\def\sc{\mathsf{c}}

\def\sq{\mathsf{q}}

\def\sw{\mathsf{w}}%

\def\sK{\mathsf{K}}

\def\sS{\mathsf{S}}

\def\sZ{\mathsf{Z}}
 
\def\sM{\mathsf{M}}

\def\sW{\mathsf{W}}

\def\trasp{\mathsf{T}}

\newcommand{\xp}{\mathbb{E}}

\def\qmo{``}
\def\qmc{''}
\def\qmcsp{'' }

\newcounter{example}[section]
\def\theexample{\thesection.\arabic{example}}
{
\medskip\noindent\\
\refstepcounter{example} {\bf Example \theexample\hspace{2pt}}
-\hspace{2pt}}{%
 \begin{flushright}
 {\footnotesize $\square$}
 \end{flushright}
}

\newenvironment{proof2}{{\bfseries Proof ---}
\parindent=0pt}{\parfillskip=0pt\hfil$\square\square$\par\bigbreak}

\newenvironment{proofdot}{\noindent{\bf Proof.} \it}{\hfill
$\Box\Box$ \bigskip\newline}

\title{\LARGE\bf Multivariate Markov-Switching models \\ and tail risk interdependence}

\author{M. Bernardi\footnote{ MEMOTEF, Sapienza University of Rome}, A. Maruotti\footnote{Southampton Statistical Sciences Research Institute \& School of Mathematics, University of Southampton}  and L. Petrella\footnote{Corresponding author:  MEMOTEF, Sapienza University of Rome, Via del Castro Laurenziano, 9, 00161 ROME, \textrm{Tel.:} +39.06.49766972, \textrm{email address:} \textrm{lea.petrella@uniroma1.it.}}}

\begin{document}

\maketitle

\date{}

\begin{abstract}
Markov Switching models are often used to analyse financial returns because of their ability to capture frequently observed stylised facts. In this paper we consider a multivariate Student-t version of the model as a viable alternative to the usual multivariate Gaussian distribution, providing a natural robust extension that accounts for heavy--tails and time varying non--linear correlations. Moreover, these modelling assumptions allow us to capture extreme tail co--movements which are of fundamental importance to assess the underlying dependence structure of asset returns during extreme events such as financial crisis. In order to capture risk interdependence among several multiple connected market participants which may experience contemporaneously distress instances, we provide new multiple risk interdependence measures generalising the Conditional Value-at-Risk (CoVaR) and the Conditional Expected Shortfall (CoES) of Adrian and Brunnermeier \cite{adrian_brunnermeier.2011}.  Those measures are analytically evaluated on the predictive distribution of the models in order to provide a forward--looking risk quantification. Application on a set of U.S. banks is considered to show that the right specification of the model conditional distribution along with a multiple risk interdependence measure may help to better understand how the overall risk is shared among institutions.\newline\newline   
%
%
\noindent \textsc{Keywords}: Markov Switching, tail risk interdependence, risk measures, CoVaR
\vspace{0.25cm}\\
%
\end{abstract}

\newpage

\section{Introduction}
\label{sec:intro}
%
The recent global financial crisis originated by the U.S. subprime mortgage bubble burst of August 2007 and the consequent downturn in economic activity leading to the 2008--2012 global recession, highlighted the strong negative impact of large scale collapse of financial institutions on other banks as well as on the real economy. After the failure of Bearn Stearns hedge funds on August 5th, 2007, the threat of total collapse of large financial institutions, and the consequent downturns in stock markets around the world, caused a worldwide increase in financial market volatility and a sudden tightening of the liquidity conditions. The spillover effect of a downturn in the financial system has been advocated as the main reason for massive public interventions and bailouts of distressed banks. Bank failures cause direct effects on the real economy because of their linkage to the manufactory industry through the credit mechanism as well as the significant role they play as financial intermediaries on the monetary transmission channel. Notwithstanding, before the 2007--2008 crisis, banking regulation and risk capital allocation was based on individual risk measures such as the Value-at-Risk (VaR). Unfortunately, such risk measure fails to consider the institution as part of a system which might itself experience instability and spread new sources of systemic risk. Recently, Adrian and Brunnermeier \cite{adrian_brunnermeier.2011} introduced the so called Conditional Value-at-Risk (CoVaR), as one of the possible measures of systemic risk, which is defined as the overall Value at Risk (VaR) of an institution conditional on another institution being under distress. In this way the CoVaR not only captures the overall risk embedded in each institution, but also reflects individual contributions to the systemic risk, capturing extreme tail co--movements, see for example Bernardi \textit{et al.} \cite{bernardi_etal.2013}, Girardi and Erg\"{u}n \cite{girardi_ergun.2013} and Castro and Ferrari \cite{castro_ferrari.2013}. However, recent financial crisis are characterised by the contemporaneous distress of several institutions emphasising the difficulty to accurately measure marginal contributions to overall risk of an institution taken in isolation. The spillover effect of a financial downturn may propagate through other institutions being in distress at the same time. As a consequence, new overall risk measures that account for contemporaneous multiple distress as conditioning events should be delivered.\newline
\indent In this paper, we develop a multivariate model--based approach to measure the dynamic evolution of tail risk interdependence accounting for the well known characteristics of financial time series. To achieve this goal, we consider the class of multivariate Markov Switching Models (MSMs), see e.g. Ang and Bekaert \cite{ang_bekaert.2004}, Bulla \textit{et al.} \cite{bulla_etal.2011}. Multivariate MSMs may be recognised as a challenging and promising approach in modelling financial time series, mainly because of their ability to reproduce some of the most important stylised facts and to account for nonlinearity and persistence of the visited states (see e.g. Harris and K\"{u}\c{c}\"{u}k\"{o}zmen \cite{harris_kucukukozmen.2001} and Gettinby \textit{et al.} \cite{gettinby_etal.2004}). Those features are crucial aspects in market return analysis and risk modelling. Markov Switching models has been intensively used in literature see e.g. Ryd\'en \textit{et al.} \cite{ryden_etal.1998}, Ang and Bekaert \cite{ang_bekaert.2004}, Bulla and Bulla \cite{bulla_and_bulla.2006}, Bulla \textit{et al.} \cite{bulla_etal.2011}, Hamilton \cite{hamilton.1989} Hamilton \cite{hamilton.1990}. Moreover in their papers, Amisano and Geweke \cite{amisano_geweke.2011} and Geweke and Amisano \cite{geweke_amisano.2010} showed that MSMs outperform their competitors in predicting daily returns of financial time series, especially during volatile episodes.\newline
\indent Throughout this paper, we consider both multivariate Gaussian and Student--t assumptions to model the conditional distribution of the observations. The choice between the two distributions is empirically tested on real data although the multivariate Normal specification may be considered a starting point for our analysis. In fact, it is well known that multivariate Gaussian distributions have some deficiencies in particular when financial time series and risk assessment problems are considered. Firstly, they rely on the assumption that each marginal follows a Normal distribution, which is often unrealistic for daily returns, exhibiting extreme realizations (see e.g. Bulla \cite{bulla.2011}). More importantly, as documented in Embrechts \textit{et al.} \cite{embrechts_etal.2000}, non--linearities or strong deviation from the elliptical and symmetric distribution for the underlying observed process may affect the association measures and, as a consequence, the covariance will not capture the complete dependence anymore. Moreover, the tails of the multivariate Gaussian distribution lead to independent extremes, resulting in potentially underestimation of the probabilities of simultaneous occurrence of rare events (see e.g. \v{C}i\v{z}ek \textit{et al.} \cite{cizek_etal.2011}, Demarta and McNeil \cite{demarta_mcneal.2005} and McNeil \textit{et al.} \cite{mcneil_etal.2005}). 
The latter eventuality may cause incorrect assessment of economic risks which has played a key role in the ongoing international financial crisis. 
To overcome all the above mentioned limitations of the multivariate Gaussian distribution we consider multivariate Student--t MSMs as a natural robust extension.\newline 
\indent Once the modelling framework has been properly set, we develop a consistent approach to measure tail risk interdependence among institutions. We introduce Multiple--CoVaR and Multiple--CoES risk measures extending and improving the Adrian and Brunnermeier's CoVaR and CoES. The proposed risk measures aim to capture interconnections among multiple connecting market participants which is particularly relevant during periods of financial market instability, when several institutions may contemporaneously experience distress instances. Those measures are analytically evaluated on the MSM predictive distribution in order to provide a forward--looking risk quantification. In particular, since the predictive distribution of MSMs is a finite mixture of the Markovian component densities we provide analytical formula to compute Multiple--CoVaR and Multiple--CoES assuming that the conditioning financial event refers to a set of institutions being under distress, under both multivariate Gaussian and Student--t assumptions. Building risk measures upon multivariate MSMs allows to differentiate institutions' tail risk exposure depending on the state of the economy identified by the latent Markovian process. Moreover, the proposed framework enables us to capture the time varying exposure of individual institutions with respect to other institutions' VaR levels.\newline
\indent The strategy developed in this paper to assess institutions' risk contributions relies on the Multiple--$\Delta$CoVaR and Multiple--$\Delta$CoES. Those marginal contributions are measured as the difference between the Multiple--CoVaR (Multiple--CoES) of each institution $j$ conditional on a given set of different institutions being under distress and the Multiple--CoVaR (Multiple--CoES) of institution $j$ evaluated when the same set of conditioning institutions are at their normal state, identified as the median state. Whenever $j$ is assumed to be the market index, the risk measures introduced throughout the paper identify the systemic risk.\newline 
\indent In evaluating multiple risk measures, different conditioning distress events can be considered. In order to attribute the overall risk to each participant we apply the Shapley value methodology. The idea behind the Shapley value, initially proposed by Shapley \cite{shapley.1953} in the field of cooperative games, has been previously considered by Tarashev \textit{et al.} \cite{tarashev_etal.2010} and Cao \cite{cao.2013} in the field of systemic risk attribution.
%
%
%
The resulting Shapley value risk measure is additive which ensures that the final risk allocation is efficient. This property allows to overcome the deficiency of the standard CoVaR definition of Adrian and Brunnermeier \cite{adrian_brunnermeier.2011} for which the sum of individual contributions does not equal the total risk measure, providing misleading informations for policy purposes.\newline
\indent The developed methodology is applied to five major US banks belonging to the Standard and Poor's 500 index in order to assess individual institutions' marginal contribution to the systemic risk. Comparing the assumption on the MSM component density we observe that the Student--t distribution is preferred to the Gaussian one and this supports the use of fat-tails distributions. Our multivariate Student--t MSM is able to distinguish and cluster time periods corresponding to different risk--returns profiles and to model the 
persistence of visited states. Moreover, we provide the dynamic evolution of the total systemic risk as well as the marginal contribution of each considered bank. Concerning the total risk, we find that the overall systemic risk during the 2007--2009 financial crisis is larger than the total systemic risk during the European sovereign debt crisis at its peak. Our main empirical result suggests that the marginal contribution to the systemic risk of individual banks varies through time and particularly during periods of financial instability it changes dramatically, both in order of importance and in levels. As a further empirical output, we notice that comparing the institutions' risk contribution ordering we observe that the two measures provide slightly different results. This discordance can be probably ascribable to the 
known coherence deficiency of the VaR that rebounds on the CoVaR.\newline
\indent The remainder of the paper is organised as follows. Section \ref{sec:hmm_models} introduces the multivariate Gaussian and Student--t MSMs and provides the estimation methodology. Section \ref{sec:risk_measures} provides some useful results concerning the marginal and conditional distributions of multivariate mixtures and the new risk measures we introduce, while Section \ref{sec:systemic_risk_contribution} details the Shapley value methodology to evaluate individual contributions to systemic risk. Section \ref{sec:empirical_analysis} presents results based on an illustrative basket of five major US banks belonging to the S\&P500 composite index. Section \ref{sec:conclusion} concludes.
\section{Markov switching models}
\label{sec:hmm_models}
%
We provide a brief introduction to MSMs setup, focusing on multivariate Gaussian and Sudent--t distribution assumption. For an up to date review of MSMs see e.g. Capp\'e \textit{et al.} \cite{cappe_etal.2005}, Zucchini and MacDonald \cite{zucchini_macdonald.2009} and Dymarski \cite{dymarski.2011}.
%
\subsection{Model setup}
\label{sec:hmm_setup}
%
In this section we illustrate the formulation of the MSM for multivariate continuous response variables. Let $\left\{\bY_t,t=1,\dots,T\right\}$ denote a sequence of multivariate observations, where $\bY_t = \{\sY_{1,t},\sY_{2,t},\dots,\sY_{\sp,T}\}\in {\mathbb R}^\sp$, while $\left\{\sS_t,t=1,\dots,T\right\}$ denotes a Markov chain defined on the state space $\{1,2,\dots,L\}$. A MSM is a stochastic process consisting of two parts: the underlying unobserved process $\left\{\sS_t\right\}$, fulfilling the Markov property, i.e.
\begin{equation}
\mathbb{P}\left(\sS_t=\mathsf{s}_t\mid \sS_1=\mathsf{s}_{1},\sS_2=\mathsf{s}_{2},\dots,\sS_{t-1}=\mathsf{s}_{t-1}\right)
=\mathbb{P}\left(\sS_t=\mathsf{s}_t\mid \sS_{t-1}=\mathsf{s}_{t-1}\right),\nonumber
\end{equation}
and the state--dependent observation process $\{\bY_t\}$ for which the conditional independence property, i.e.
\begin{equation}
f\left(\bY_t=\by_t\mid \bY_1=\by_1,\dots,\bY_{t-1}=\by_{t-1},\sS_1=\mathsf{s}_1,\dots,\sS_t=\mathsf{s}_t\right)=f\left(\bY_t=\by_t\mid \sS_t=\mathsf{s}_t\right),\nonumber
\end{equation}
holds, where $f\left(\cdot\right)$ is a generic probability density function.\newline
\indent The literature on MSMs  for continuous data  is dominated by Gaussian MSMs (Hamilton \cite{hamilton.1989}, Ryd\'en \textit{et al.} \cite{ryden_etal.1998}, Bialkowski \cite{bialkowski.2003}, Bartolucci and Farcomeni \cite{bartolucci_farcomeni.2010}), with few exceptions (Bartolucci and Farcomeni \cite{bartolucci_farcomeni.2009} and Lagona and Picone \cite{lagona_picone.2013}). Under the Gaussian assumption, the state--specific distribution of $\bY_t$ is given by 
\begin{equation}
\label{pdf_gaussian}
\bY_t\mid\sS_t=\mathsf{s}_t\sim\mathcal{N}_\sp\left(\bmu_{\mathsf{s}_t},\bSigma_{\mathsf{s}_t}\right)
\end{equation} 
where $\mathcal{N}_\sp\left(\bmu_{\mathsf{s}_t},\bSigma_{\mathsf{s}_t}\right)$ denotes the multivariate Gaussian distribution with mean $\bmu_{\mathsf{s}_t}$ and covariance matrix $\bSigma_{\mathsf{s}_t}$.\newline
%
%
\indent Time series models for financial data should account for several well known departures from normality such as heavy--tails, robustness to outliers, and the ability of capturing extreme events.
Those reasons motivate our choice of the multivariate Student--t assumption for the MSM:
%
%
%
\begin{equation}
\label{eq:pdf_student}
\bY_t\mid\sS_t=\mathsf{s}_t\sim\mathcal{T}_{\sp}\left(\bmu_{\mathsf{s}_t},\bSigma_{\mathsf{s}_t},\nu_{\mathsf{s}_t}\right)
\end{equation} 
where $\mathcal{T}_\sp\left(\bmu_{\mathsf{s}_t},\bSigma_{\mathsf{s}_t},\nu_{\mathsf{s}_t}\right)$ denotes the multivariate Student--t distribution with mean $\bmu_{\mathsf{s}_t}$, scale matrix $\bSigma_{\mathsf{s}_t}$ and degrees of freedom equalt to $\nu_{\mathsf{s}_t}$. 
%
%
%
As $\nu_{\mathsf{s}_t}$ tends to infinity, the distribution in equation (\ref{eq:pdf_student}) approaches the Gaussian distribution, with mean $\bmu_{\mathsf{s}_t}$ and variance--covariance matrix $\bSigma_{\mathsf{s}_t}$. Hence the parameter $\nu_{\mathsf{s}_t}$ may be viewed as a robustness tuning parameter, which need to be estimated along with all other model parameters.\newline
\indent For further technical purposes, we remind that the multivariate Student-t distribution can be expressed as a scale mixture of multivariate Gaussian distributions
\begin{equation}
\label{eq:student_scale_mixture}
\bY_t\mid\left(\sS_t=\mathsf{s}_t, \sW_t=\sw_t\right)\sim
\mathcal{N}_\sp\left(\bmu_{\mathsf{s}_t},\frac{\bSigma_{\mathsf{s}_t}}{\sw_t}\right),
\end{equation}
where the mixing variables $\left\{\sW_t,t=1,2,\dots,T\right\}$ are independent and identically distributed random variables having distribution
\begin{equation} 
\label{eq:student_MS_augmenting_variable}
\sW_t\mid\sS_t=\mathsf{s}_t\sim\mathcal{G}\left(\frac{1}{2}\nu_{\mathsf{s}_t},\frac{1}{2}\nu_{\mathsf{s}_t}\right)
\end{equation}
with $\mathcal{G}\left(\alpha,\beta\right)$ denoting the Gamma distribution with parameters $\alpha>0$ and $\beta>0$, see e.g. Kotz and Nadarajah \cite{kotz_nadarajah.2004}.\newline
%
%
%
%
%
%
\indent In order to complete the model setup, we need to specify the Markov chain which determines the hidden state at each time point $t$. To this purpose let denote with $\sq_{l,k}=\mathbb{P}\left(\sS_t=k\mid\sS_{t-1}=l\right)$, $\forall l,k\in\left\{1,2,\dots,L\right\}$ the probability that state $k$ is visited at time $t$ given that at time $t-1$ the chain was visiting state $l$. We indicate with $\delta_l=\mathbb{P}\left(\sS_1=l\right)$ the initial probability of being in state $l=\left\{1,2,\dots,L\right\}$ at time 1, and we refer to $\bQ=\{\sq_{l,k}\}$ as the transition probability matrix of the Markov chain.
%
%
\subsection{Estimation and inference}
\label{subsec:hmm_inference}
%
The MSM parameters are generally estimated using the maximum--likelihood method, see for example McLachlan and Peel \cite{mclachlan_peel.2000} and Capp\'e \textit{et al.} \cite{cappe_etal.2005}. The likelihood of a MSM can be expressed in a closed--form formula, even in a relatively general framework. Let $\boldsymbol{\theta}=\left\{\bmu_l,\bSigma_l,\nu_l,\bQ,\bdelta,l=1,2,\dots,L\right\}$ be the set of all model parameters and let $\bff\left(\by_t\right)$ denote a diagonal matrix with conditional probabilities $f\left(\bY_t=\by_t\mid \sS_t=\mathsf{s}_t\right)$ on the main diagonal, then, the likelihood of a MSM can be written as
\begin{equation}
\label{eq:MVT_HMM_loglike} 
\mathcal{L}\left(\boldsymbol{\theta}\right)=\bdelta\bff\left(\by_1\right)\bQ\bff\left(\by_2\right)\bQ\dots \bff\left(\by_{T-1}\right)\bQ{\bff}(\by_T)\mathbf{1}'.
\end{equation}
Finding the value of the parameters $\boldsymbol{\theta}$ that maximize the log--transformation of equation (\ref{eq:MVT_HMM_loglike}) under the constraints $\sum_{l=1}^L\delta_{l}=1$ and $\sum_{k=1}^\sK \sq_{l,k}=1$, is not an easy problem since (\ref{eq:MVT_HMM_loglike}) is not analytically available. Instead, it is straigthforward to find solutions of equation (\ref{eq:MVT_HMM_loglike}) using the Expectaton--Maximization (EM) algorithm of Dempster \textit{et al.} \cite{dempster_etal.1977}. Hereafter we focus on the EM algorithm which has been previously applied to the case of finite mixtures of univariate Student--t distributions by Peel and McLachlan \cite{peel_mclachlan.2000}.\newline
\indent For the purpose of application of the EM algorithm the vector of observations $\by_t,t=1,2,\dots,T$ is regarded as being incomplete. Following the implementation described in Peel and McLachlan \cite{peel_mclachlan.2000} in a finite mixture context, two missing data structures are consequently introduced.
%
%
The first one is related to the unobservable Markovian states, i.e. $\bz_{t}=\left(\sz_{t,1},\sz_{t,2},\dots,\sz_{t,L}\right)$ and $\bz\bz_{t}=\left(\sz\sz_{t,1,1},\sz\sz_{t,1,2},\dots,\sz\sz_{t,l,k},\dots,\sz\sz_{t,L,L}\right)$ defined as
\begin{eqnarray}
\sz_{t,l}&=&\left\{\begin{array}{cc}1 &  \text{if}\quad \sS_t=l\\ 0 & \text{otherwise}\end{array}\right.\nonumber\\
\sz\sz_{t,l,k}&=&\left\{\begin{array}{cc}1 & \text{if}\quad \sS_{t-1}=l, \sS_{t}=k\\ 0 & \text{otherwise.}\end{array}\right.\nonumber
\end{eqnarray}
%
%
The second type of missing data structure is $\sw_t$, $\forall t=1,2,\dots,T$ relies to the scale mixture representation in equations (\ref{eq:student_scale_mixture})-(\ref{eq:student_MS_augmenting_variable}) which are assumed to be conditionally independent given the component labels $\sz_{l,t},l=1,2,\dots,L,\forall t=1,2,\dots,T$.\newline
\indent Augmenting the observations $\left\{\bY_t,t=1,2,\dots,T\right\}$ with the latent variables $\left\{\sw_t,\sz_{tl},\sz\sz_{t,l,k},t=1,2,\dots,T; l = 1,\dots,L\right\}$ 
gives the following complete--data log--likelihood:
%
\begin{eqnarray}
\label{eq:MST_complete_loglike}
\log\mathcal{L}_{\sc}\left(\boldsymbol{\theta}\right)
&=&\sum_{l=1}^L\sz_{1,l} \log\left(\delta_l\right)\nonumber\\ 
&+& \sum_{l=1}^L\sum_{k=1}^L\sum_{t=1}^T\sz\sz_{t,l,k}\log\left(\sq_{l,k}\right) \nonumber\\ 
&+&\sum_{l=1}^L\sum_{t=1}^T\sz_{t,l}\left\{-\frac{p}{2}\log\left(2\pi\right)-\frac{1}{2}\log\vert\bSigma_l\vert-\frac{\sw_t}{2}\left(\by_t-\bmu_{l}\right)^\trasp\bSigma_{l}^{-1}\left(\by_t-\bmu_{l}\right)\right\}\nonumber\\
&+&\sum_{l=1}^L\sum_{t=1}^T\sz_{t,l}\left\{\frac{\nu_l}{2}\log\left(\frac{\nu_l}{2}\right)-\log\Gamma\left(\frac{\nu_l}{2}\right)\right\}\nonumber\\
&+&\sum_{l=1}^L\sum_{t=1}^T\sz_{t,l}\left\{\frac{\nu_l}{2}\left(\log (\sw_t)-\sw_t\right)+\left(\frac{p}{2}-1\right)\log\left(\sw_t\right)\right\}.
\end{eqnarray}
%
\noindent The EM algorithm consists of two major steps, one for expectation (E--step) and one for maximization (M--step), see McLachlan and Krishnan \cite{mclachlan_krishnan.2008}. 
At the $\left(m+1\right)$--th iteration the EM algorithm proceeds as follows:
\begin{itemize}
\item[\textbf{E--step:}] computes the conditional expectation of the complete--data log--likelihood (\ref{eq:MST_complete_loglike}) given the observed data $\left\{\by_t\right\}_{t}^T$ and the $m$--th iteration $\boldsymbol{\theta}^{\left(m\right)}$ parameters estimates
\begin{equation}
\mathcal{Q}\left(\boldsymbol{\theta},\boldsymbol{\theta}^{\left(m\right)}\right)=\xp_{\boldsymbol{\theta}^{\left(m\right)}}
\left[\log\mathcal{L}_{\sc}\left(\boldsymbol{\theta}\right)\mid\left\{\by_t\right\}_{t}^T\right]
\label{eq:estep}
\end{equation}
\item[\textbf{M--step:}] choose $\boldsymbol{\theta}^{\left(m+1\right)}$ by maximising (\ref{eq:estep}) with respect to $\boldsymbol{\theta}$
\begin{equation}
\boldsymbol{\theta}^{\left(m+1\right)}=\arg\max_{\boldsymbol{\theta}}\mathcal{Q}\left(\boldsymbol{\theta},\boldsymbol{\theta}^{\left(m\right)}\right).
\end{equation}
\end{itemize}
One nice feature of the EM algorithm is that the solution of the M--step exists analytically for Gaussian and Student--t MSMs, with the exception of the degrees--of--freedom $\nu_l,l=1,2,\dots,L$. 
One possible solution for estimating $\nu_l$ is to adopt the approximation provided in Shoham \cite{shoham.2002} for mixtures of Student--t distributions. In Appendix \ref{sec:appendix_A} we sketch the corresponding EM algorithm.
%
\section{Risk measures}
\label{sec:risk_measures}
%
As discussed in the Introduction, one of the main contribution of this paper is a model--based approach to quantify interdependence tail risk. Assessing financial risks requires the appropriate definition of risk measures accounting for potential spillover effects among institutions belonging to a given financial market. After the 2007--2008 global financial crisis a large stream of literature has been devoted to this topic: see for example the Marginal Expected Shortfall risk measure (MES) of Acharya \textit{et al.} \cite{acharya_etal.2010}, the Systemic--RISK measure (SRISK) jointly proposed by Brownlees and Engle \cite{brownlees_engle.2012} and Acharya \textit{et al.} \cite{acharya_etal.2012}, for a portfolio approach to measure the total overall system--wide risk. Acharya and Richardson \cite{acharya_richardson.2009}, Huang \textit{et al.} \cite{huang_etal.2012}, Billio \textit{et al.} \cite{billio_etal.2012} instead measure the marginal contribution of individual institutions to the systemic risk. In this paper we follow this latter stream of literature focusing in particular on the recent work of Adrian and Brunnermeier \cite{adrian_brunnermeier.2011} who introduce the CoVaR approach to measure overall risk contributions of individual institutions. The original definition of CoVaR at $\tau$--level, i.e. ${\rm CoVaR}_{i\vert j}^{\tau}$, considers two different institutions $i$ and $j$, 
such that
\begin{equation}
\mathbb{P}\left(\sY_i\leq {\rm CoVaR}_{i\vert j}^{\tau}\mid \sY_j={\rm VaR}_j^{\tau}\right)=\tau
\end{equation}
where $\sY_i$ and $\sY_j$ denote the institution $i$ and $j$ returns and ${\rm VaR}_j^{\tau}$ denotes the \qmo univariate marginal\qmcsp Value--at--Risk of asset $j$. If $i$ coincides with the whole financial system, the CoVaR becomes the VaR of the financial system conditional on a single institution $j$ being in financial distress and represents the basis to understand how systemic risk shares among institutions.\newline 
\indent Recent financial crisis are characterised by the contemporaneous default of several institutions underlying the exigence of risk measures accounting for joint occurrence of extreme losses. For this reason in Subsection \ref{subsec:covar_coes} we introduce two new systemic risk measures; the first one is the Multiple--CoVaR, extending the CoVaR of Adrian and Brunnermeier \cite{adrian_brunnermeier.2011} to the case where more than a single institution experiences distress instances. 
The second one is the Multiple--CoES which generalizes the $\tau$--level ${\rm CoES}_{i\vert j}^{\tau}$ of Adrian and Brunnermeier \cite{adrian_brunnermeier.2011} i.e.
\begin{equation}
{\rm CoES}_{i\vert j}^{\tau}\equiv\mathbb{E}\left(\sY_i\mid\sY_i\leq\hat{\sy}_{i}^{\tau},\sY_j={\rm ES}_j^{\tau}\right)
\end{equation}
where ${\rm ES}_j^{\tau}$ is the univariate marginal expected shortfall of asset $j$.
By definition, the Multiple--CoVaR and Multiple--CoES rely on the conditional and marginal return distributions. In what follows we embed the proposed risk measures in the MS framework defined in the previous Section \ref{sec:hmm_setup}. In particular, we focus on the \qmo predictive\qmcsp distribution of the Markov--Switching model and we provide analytical expressions for marginal and conditional predictive distributions which allows us to give explicit formulae for Multiple--CoVaR and Multiple--CoES   under the Gaussian and Student-t assumptions. 
%
%
%

%
\subsection{Preliminary results}
\label{sec:preliminary results}
%
%
In this subsection we firstly recall a known result on the $h$--steps ahead \qmo predictive\qmcsp distribution for the MSM. We then use such a result to prove two theorems concerning its marginals and conditionals distributions under the multivariate Gaussian and multivariate t assumption for the component specific density.\newline\newline
%
%
%
\noindent The $h$-step ahead distribution of the observed process $\mathbf{y}_{t+h}$ at time $t+h$, given information up to time $t$, $\mathcal{F}_t$, is a finite mixture of component specific predictive distributions, (see e.g. Zucchini and MacDonald \cite{zucchini_macdonald.2009})
\begin{eqnarray}
%
p\left(\mathbf{y}_{t+h}\mid\mathcal{F}_t\right)
=\sum_{l=1}^L\pi_{l}^{\left(h\right)}f\left(\mathbf{y}_{t+h}\mid S_{t+h}=l\right)
\label{eq:hmm_predictive_general}
\end{eqnarray}
with mixing weights
\begin{eqnarray}
\label{eq:predictive_mix_weights}
\pi_{l}^{\left(h\right)}
=\sum_{j=1}^L\mathbf{Q}_{j,l}^{h}\mathbb{P}\left(S_{t}=j\mid\mathcal{F}_t\right),
\end{eqnarray}
where $\mathbf{Q}_{j,l}^{h}$ is the $\left(j,l\right)$-th entry of the Markovian transition matrix $\mathbf{Q}$ to the power $h$. 
%
%
%
%
Under multivariate Gaussian and Student--t assumptions on $f\left(\mathbf{y}_{t+h}\mid S_{t+h}\right)$ we have the following results.
%
%
\begin{theorem}[\textbf{Conditional and marginal distributions for Multivariate Gaussian Mixtures}]
\label{th:GaussMix_cond_marg_dist} 
Let $\bY$ be a $\sp$-dimensional Gaussian mixture, i.e. $\bY\sim\sum_{l=1}^L\eta_l \mathcal{N}_\sp\left(\by\mid\bmu_l,\bSigma_l\right)$, and assume $\bY$ partitioned into $\bY=\left[\bY_1^\trasp,\bY_2^\trasp\right]^\trasp$, where $\bY_1$ and $\bY_2$ are of dimension $dim\left(\bY_1\right)=\sp_1$ and $\text{dim}\left(\bY_2\right)=\sp_2=\sp-\sp_1$, respectively. The mean vectors $\bmu_l$ and the variance-covariance matrices $\bSigma_l$ for $l=1,2,\dots,L$ are partitioned accordingly to $\bmu_l=\left[\bmu_l^{1\trasp},\bmu_l^{2\trasp}\right]^\trasp$ and $\bSigma_l=\left[\begin{array}{cc}\bSigma_l^{\left(1,1\right)} & \bSigma_l^{\left(1,2\right)} \\ \bSigma_l^{\left(2,1\right)} & \bSigma_l^{\left(2,2\right)}\end{array}\right]$, respectively, where $\bSigma_l^{(2,2)}$ is a $\sp_2\times \sp_2$ positive definite matrix. Then:
\begin{itemize}
\item[(i)] the marginal distribution of $\bY_2$ is
\begin{eqnarray*}
f_\mathsf{\bY_2}\left(\mathsf{\by_2}\right)
=\sum_{l=1}^L\eta_l\mathcal{N}_{p_2}\left(\by_2\mid\bmu_l^{2},\bSigma_l^{(2,2)}\right)
\end{eqnarray*}
\item[(ii)] the conditional distribution of $\bY_1$ given $\bY_2=\by_2$ is
\begin{equation*}
f_{\bY_1}\left(\by_1\mid\bY_2=\by_2\right)=\sum_{l=1}^L\tilde{\eta}_l\left(\by_2\right)\mathcal{N}_{\sp_1}\left(\by_1\mid\bmu_l^{1\vert 2}\left(\by_2\right),\bSigma_l^{1\vert 2}\right),
\end{equation*}
where the mixing weights have the following expression
\begin{eqnarray}
\tilde{\eta}_l\left(\by_2\right)=
\frac{\eta_l\mathcal{N}_{p_2}\left(\by_2\mid\bmu^2_l,\bSigma^{(2,2)}_l\right)}{\sum_{l=1}^L\eta_l\mathcal{N}_{p_2}\left(\by_2\mid\bmu_l^{2},\bSigma^{(2,2)}_l\right)},
\label{eq:multivGaussMix_Cond_weights}
\end{eqnarray}
and the conditional moments of each components are 
\begin{eqnarray}
\bmu_l^{1\vert 2}\left(\by_2\right)
&=&\bmu_l^1+\bSigma_l^{\left(1,2\right)}\bSigma_l^{\left(2,2\right)^{-1}}\left(\by_2-\bmu_l^{2}\right)\label{eq:gauss_mix_cond_mean}\\ 
\bSigma_{l}^{1\vert 2}&=&\bSigma_{l}^{\left(1,1\right)}-\bSigma_{l}^{\left(1,2\right)}\bSigma_{l}^{\left(2,2\right)^{-1}}\bSigma_{l}^{\left(2,1\right)}, \quad\forall l=1,2,\dots,L.
\label{eq:gauss_mix_cond_var}
\end{eqnarray}
\end{itemize}
\end{theorem}

\begin{proofdot} 
\begin{itemize}
\item[(i)] The first result follows from standard integration.
\item[(ii)] Concerning the second result, applying result (i) we can factorize the joint density in the following way:
\begin{eqnarray*}
f_{\bY_1\vert\bY_2=\by_2}\left(\by_2\right)&=&\frac{f_{\bY_1,\bY_2}\left(\by_1,\by_2\right)}{f_{\bY_2}\left(\by_2\right)}\\
&=&\frac{\sum_{l=1}^L\eta_l\mathcal{N}_{p_2}\left(\by_2\mid\bmu^2_l,\bSigma^2_l\right)\mathcal{N}_{p_1}\left(\bY_1\mid\bY_2=\by_2,\bmu^{1\vert 2}_l\left(\by_2\right),\bSigma^{1\vert 2}_l\right)}{\sum_{l=1}^L\eta_l\mathcal{N}_{p_2}\left(\by_2\mid\bmu^2_l,\bSigma^{(2,2)}_l\right)}
\end{eqnarray*}
which is a mixture of multivariate Gaussian distributions with mixing proportions defined as in equation (\ref{eq:multivGaussMix_Cond_weights}).
\end{itemize}
\end{proofdot}
\begin{theorem}[\textbf{Conditional and marginal distributions for Multivariate Student-t Mixtures}]
\label{th:StudentMix_cond_marg_dist} 
Let $\bY$ be a $p$-dimensional Student--t mixture, i.e. $\bY\sim\sum_{l=1}^L\eta_l \mathcal{T}_p\left(\by\mid\bmu_l,\bSigma_l,\nu_l\right)$, and assume $\bY$ partitioned into $\bY=\left[\bY_1^\trasp,\bY_2^\trasp\right]^\trasp$, where $\bY_1$ and $\bY_2$ are of dimension $dim\left(\bY_1\right)=\sp_1$ and $\text{dim}\left(\bY_2\right)=\sp_2=\sp-\sp_1$, respectively. The location parameter vectors $\bmu_l$ and the scale matrices $\bSigma_l$ for $l=1,2,\dots,L$ are partitioned accordingly to $\bmu_l=\left[\bmu_l^{1\trasp},\bmu_l^{2\trasp}\right]^\trasp$ and $\bSigma_l=\left[\begin{array}{cc}\bSigma_l^{\left(1,1\right)} & \bSigma_l^{\left(1,2\right)} \\ \bSigma_l^{\left(2,1\right)} & \bSigma_l^{\left(2,2\right)}\end{array}\right]$, respectively, where $\bSigma_l^{(2,2)}$ is a $\sp_2\times \sp_2$ positive definite matrix. Then:
\begin{itemize}
\item[(i)] the marginal distribution of $\bY_2$ is
\begin{eqnarray*}
f_\mathsf{\bY_2}\left(\mathsf{\by_2}\right)
=\sum_{l=1}^L\eta_l\mathcal{T}_{p_2}\left(\by_2\mid\bmu_l^{2},\bSigma_l^{(2,2)},\nu_l\right)
\end{eqnarray*}
\item[(ii)] the conditional distribution of $\bY_1$ given $\bY_2=\by_2$ is
\begin{equation*}
f_\mathsf{\bY_1}\left(\mathsf{\by_1\mid\bY_2=\by_2}\right)
=\sum_{l=1}^L\tilde{\eta}_l\left(\by_2\right)\mathcal{T}_{p_1}\left(\by_1\mid\bmu_l^{1|2}\left(\by_2\right),\bSigma_l^*,p_1+\nu_l\right),
\end{equation*}
where $\bSigma^*=\left[1+\frac{1}{\nu_l}\left(\by_2-\bmu_l^2\right)^{\trasp}\bSigma_l^{(2,2)^{-1}}\left(\by_2-\bmu_l^2\right)\right]\bSigma_l^{1\vert 2}\frac{\nu_l}{p_1+\nu_l}$, the mixing weights have the following expression
\begin{eqnarray}
\tilde{\eta}_l\left(\by_2\right)=
\frac{\eta_l\mathcal{T}_{p_2}\left(\by_2\mid\bmu^2_l,\bSigma^{(2,2)}_l,\nu_l\right)}{\sum_{l=1}^L\eta_l\mathcal{T}_{p_2}\left(\by_2\mid\bmu_l^{2},\bSigma^{(2,2)}_l,\nu_l\right)},
\label{eq:multivStudentMix_Cond_weights}
\end{eqnarray}
and $\bmu_l^{1|2}$ and $\bSigma_{l}^{1|2}$ are defined as in equations (\ref{eq:gauss_mix_cond_mean}) and (\ref{eq:gauss_mix_cond_var}) respectively.
\end{itemize}
\end{theorem}

\begin{proofdot} 
\begin{itemize}
\item[(i)] The first result follows from standard integration, using the result for the marginal distribution of multivariate Student-t distributions provided by Sutradhar \cite{sutradhar.1984}.
\item[(ii)] Concerning the second part, we can factorize the joint Student-t density as in Theorem \ref{th:GaussMix_cond_marg_dist} and using the result for the conditional distribution of multivariate Student--t distributions provided by Sutradhar \cite{sutradhar.1984}, the results follows immediately.
\end{itemize}
\end{proofdot}
%
%
In what follows we give results on univariate VaR and ES measures usefull to introduce our generalization of CoVaR and CoES risk measures.   
%
\subsection{Marginal risk measures: VaR and ES}
\label{subsec:var_es}
%
The Value--at--Risk for a risky asset at a given confidence level $\tau$ is the $\left(1-\tau\right)-$quantile of the distribution of the asset return, and measures the minimum loss that can occur in the $\left(1-\tau\right)\times100\%$ of worst cases. When the random variable $\sY$ has an absolutely continuous density function $f_{\sY}\left(\sy\right)$ with cumulative density function $F_{\sY}\left(\sy\right)$, the VaR can be calculated by ${\rm VaR}_{\tau}\left(\sY\right)\equiv\hat{\sy}^{\tau}=F_{\sY}^{-1}\left(1-\tau\right)$. If the distribution of $\sY$ is a Gaussian or Student--t mixture it is straightforward to evaluate the VaR as the distribution's quantile. In his paper, Artzner \cite{artzner_etal.1999}, shows that VaR suffers for the lack of subadditivity property, and does not take into account for the benefits deriving from diversification. In principle, well diversified portfolios of risky assets should be less risky than non-diversified ones. To overcome this problem, Acerbi \cite{acerbi.2002} introduced the so-called ``\textit{spectral risk measures}'', and, among those, the Expected Shortfall (ES). The ES is a coherent risk measure (see Acerbi and Tasche \cite{acerbi_tasche.2002}) that can be defined as the expected value of $\mathsf{Y}$ truncated below the VaR. It can be calculated as the Tail Conditional Expectation (TCE) of $\sY$ conditioned at its VaR level, i.e. the expected value of $\sY$ conditional on being below a given threshold $\hat{\sy}$, $\text{ES}_{\tau}\left(\sY\right)\equiv\text{TCE}_{\sY}\left(\hat{\sy}^{\tau}\right)$. When compared to the VaR risk measure, the ES provides a more conservative measure of risk for the same degree of confidence level, and it provides an effective tool for analysing the tail of the distribution.
%
%
%
%
Bernardi \cite{bernardi.2013} shows that, under the Gaussian mixture distribution $\text{TCE}_{\sY}\left(\hat{\sy}^{\tau}\right)$ is a convex linear combination of component specific TCE. In what follows we provide a similar result under the Student--t assumption.
\begin{proposition}[\textbf{TCE for univariate Student-t mixtures}] 
\label{prop:univmixstudent_TCE}
Let $\sY$ be a univariate Student--t mixture, i.e. $\sY\sim\sum_{l=1}^L\eta_l \mathsf{St}\left(\sy\mid\mu_l,\sigma_l^2,\nu_l\right)$, then the $\text{TCE}_{\sY}\left(\hat{\sy}^{\tau}\right)$ is given by
\begin{eqnarray}
{\rm TCE}_{\sY}\left(\hat{\sy}^\tau\right)
=\sum_{l=1}^L\pi_l{\rm TCE}_{\sY,l}\left(\hat{\sy}^\tau\right)\nonumber
\end{eqnarray}
where
\begin{eqnarray}
{\rm TCE}_{\sY,l}\left(\hat{\sy}^\tau\right)=\frac{\nu_l^{\frac{1}{2}\nu_l}}{2\sqrt{\pi}}\frac{\Gamma\left(\frac{1}{2}\left(\nu_l-1\right)\right)}{\Gamma\left(\frac{1}{2}\nu_l\right)}\left[\nu_l+\left(\frac{\hat{\sy}-\mu_l}{\sigma_l}\right)^2\right]^{-\frac{1}{2}\left(\nu_l-1\right)}
\end{eqnarray}
with weights $\pi_l=\eta_l\frac{F_{\sY}\left(\hat{\sy}^\tau,\mu_l,\sigma_l^2,\nu_l\right)}{\sum_{l=1}^L\eta_l F_{\sY}\left(\hat{\sy}^\tau,\mu_l,\sigma_l^2,\nu_l\right)},\,\,\,\forall l=1,2,\dots,L$, and $\sum_{l=1}^L\pi_l=1$ where $F_{\sY}\left(\hat{\sy}^\tau,\mu,\sigma^2,\nu\right)$ is the cdf of a Student--t random variable with parameters $\left(\mu,\sigma^2,\nu\right)$ evaluated at $\hat{\sy}^\tau$.
\end{proposition}
\begin{proofdot}
See the Appendix B.
\end{proofdot}
The following corollary provides formula for ES under Student--t mixtures.
%
\begin{corollary} Under the same assumptions of Proposition \ref{prop:univmixstudent_TCE} we have
\begin{eqnarray}
\label{eq:marginal_ES_final_formula}
{\rm ES}_{\tau}\left(\sY\right)
\equiv{\rm TCE}_{\sY}\left(\hat{\sy}^{\tau}\right)=\frac{1}{\tau}\sum_{l=1}^L\pi_l\left(\hat{\sy}^{\tau}\right){\rm TCE}_{\sY,l}\left(\hat{\sy}^{\tau}\right),
\end{eqnarray}
with $\pi_l\left(\hat{\sy}^{\tau}\right)=\eta_lF\left(\hat{\sy}^\tau,\mu_l,\sigma_l^2,\nu_l\right)$, $\forall l=1,2,\dots,L$.
\end{corollary}
%
%
%
%
%
%
%
\subsection{Conditional risk measures: MCoVaR and MCoES}
\label{subsec:covar_coes}
%
As discussed at the beginning of this section, CoVaR and CoES represent measures of extreme events interdependence which go beyond traditional idiosyncratic risk measures to capture potential spillover effects among institutions. CoVaR and CoES essentially represent two by two risk measures where each individual contribution is independent on other market participants' distress. 
When the main focus is to analyse the global systemic risk it is of fundamental importance to capture interconnections among multiple connecting market participants. This is more relevant during periods of financial market crisis, when several institutions may contemporaneously experience distress instances. For this reason we propose a generalisation of the Adrian and Brunnermeier's CoVaR and CoES, namely Multiple--CoVaR (MCoVaR) and Multiple--CoES (MCoES) for multivariate Markov--Switching models. A similar multiple CoVaR has been previously introduced by Cao \cite{cao.2013}.\newline\newline
%
%
%
%
Let $\mathcal{S}=\left\{1,2,\dots,\sp\right\}$ be a set of $\sp$ institutions, we assume that the conditioning event is a set of $\sd$ institutions under distress indexed by $\mathcal{J}_{\sd}=\left\{j_1,j_2,\dots,j_\sd\right\}\subset\,_\sd\mathbb{C}_{\sp-1}$, where $_\sd\mathbb{C}_{p-1}$ is the set of all possible combinations of $\sp-1$ elements of class $\sd$, with $\sd\leq\sp-1$. Moreover, assuming that institution $i\in\mathcal{S}$ with $i\notin\mathcal{J}_{\sd}$ and $\mathcal{J}_{\sn}=\overline{\mathcal{J}_\sd}$ is the set of institutions being at the \qmo normal\qmcsp state, we define the ``Multiple--CoVaR'', $\text{MCoVaR}_{i\vert\mathcal{J}_\sd}^{\tau_1\vert \tau_2}$ as follows.
\begin{definition} 
\label{def:covar_multiple}
Let $\bY=\left(\sY_1,\sY_2,\dots,\sY_i,\dots,\sY_\sp\right)$ be the vector of institution returns, then ${\rm MCoVaR}_{i\vert\mathcal{J}_\sd}^{\tau_1\vert \tau_2}\equiv{\rm CoVaR}_{\tau_1}\left(\sY_i\mid\bY_{\mathcal{J}_\sd}=\hat{\by}_{\mathcal{J}_\sd}^{\tau_2},\bY_{\mathcal{J}_\sn}=\hat{\by}_{\mathcal{J}_\sn}^{0.5}\right)$ is the Value-at-Risk of institution $i\in\mathcal{S}$, conditional on the set of institutions $\mathcal{J}_{\sd}$ being at their individual ${\rm VaR}_{\tau_2}$-level $\hat{\by}_{\mathcal{J}_\sd}^{\tau_2}=\left(\hat{\sy}_{j_1}^{\tau_2},\hat{\sy}_{j_2}^{\tau_2},\dots,\hat{\sy}_{j_\sd}^{\tau_2}\right)$ and the set of institutions $\mathcal{J}_{\sn}=\overline{\mathcal{J}_\sd}$ being at their individual ${\rm VaR}_{0.5}$-level $\hat{\by}_{\mathcal{J}_\sn}^{0.5}=\left(\hat{\sy}_{j_{\sd+1}}^{0.5},\hat{\sy}_{j_2}^{0.5},\dots,\hat{\sy}_{j_{\sp-1}}^{0.5}\right)$
i.e., ${\rm MCoVaR}_{i\vert\mathcal{J}_\sd}^{\tau_1\vert \tau_2}$ satisfies the following equation
\begin{eqnarray}
\mathbb{P}\left(\sY_i\leq {\rm MCoVaR}_{i\vert\mathcal{J}_\sd}^{\tau_1\vert \tau_2}
\,\mid \bY_{\mathcal{J}_\sd}=\hat{\by}_{\mathcal{J}_\sd}^{\tau_2},\bY_{\mathcal{J}_\sn}=\hat{\by}_{\mathcal{J}_\sn}^{0.5}\right)=\tau_1,\quad\forall i=1,2,\dots,\sp. 
\end{eqnarray}
%
%
\end{definition}
%
%
\noindent 
We remind that ${\rm MCoVaR}_{i\vert\mathcal{J}_\sd}^{\tau_1\vert \tau_2}$ is the $\tau_1$--quantile of the conditional predictive distribution $\sY_i\mid\bY_{\mathcal{J}_\sd}=\hat{\by}_{\mathcal{J}_\sd}^{\tau_2},\bY_{\mathcal{J}_\sn}=\hat{\by}_{\mathcal{J}_\sn}^{0.5}$ defined in equations (\ref{eq:hmm_predictive_general})--(\ref{eq:predictive_mix_weights}). In order to calculate it we use results stated in Teorem \ref{th:GaussMix_cond_marg_dist} for the Gaussian case and Theorem \ref{th:StudentMix_cond_marg_dist} for the Student--t case by inverting the following cdfs:
%
%
%
%
\begin{eqnarray}
&&{\rm F}_{\sY_i}\left(\sY_i\mid\bY_{\mathcal{J}_\sd}=\hat{\by}_{\mathcal{J}_\sd}^{\tau_2},\bY_{\mathcal{J}_\sn}=\hat{\by}_{\mathcal{J}_\sn}^{0.5}\right)=\nonumber\\
&&\qquad\qquad\qquad
\sum_{l=1}^L\tilde{\eta}_l\left(\hat{\by}_{\mathcal{J}_\sd}^{\tau_2},\hat{\by}_{\mathcal{J}_\sn}^{0.5}
\right){\rm F}_{\sY_i}^{l}\left(\sY_i\mid\bY_{\mathcal{J}_\sd}=\hat{\by}_{\mathcal{J}_\sd}^{\tau_2},\bY_{\mathcal{J}_\sn}=\hat{\by}_{\mathcal{J}_\sn}^{0.5}\right)
\label{eq:condmix_cdf}
\end{eqnarray}
where the component weights $\tilde{\eta}_l\left(\hat{\by}_{\mathcal{J}_\sd}^{\tau_2},\hat{\by}_{\mathcal{J}_\sn}^{0.5}\right)$ are defined as in equations (\ref{eq:multivGaussMix_Cond_weights})--(\ref{eq:multivStudentMix_Cond_weights}) depending on the assumption made on the component densities, and ${\rm F}_{\sY_i}^{l}\left(\sY_i\mid\bY_{\mathcal{J}_\sd}=\hat{\by}_{\mathcal{J}_\sd}^{\tau_2},\bY_{\mathcal{J}_\sn}=\hat{\by}_{\mathcal{J}_\sn}^{0.5}\right)$ is the $l$--th component cdf.\newline\newline 
%
%
%
%
%
The lack of subadditivity property of the VaR suggests to introduce, in addition to the CoVaR, the Conditional Expected Shortfall (CoES), defined by Adrian and Brunnermeier \cite{adrian_brunnermeier.2011} for two institutions $i$ and $j$, as the ES evaluated on the conditional distribution of $\sY_i$ given $\sY_j$. The following definition characterise the extension of CoES to the Multiple--CoES (MCoES) accounting for multiple contemporaneous distress events.
%
%
%
\begin{definition}
Let $\bY=\left(\sY_1,\sY_2,\dots,\sY_\sp\right)$ be the vector of institution returns, then the 
\begin{equation}
{\rm MCoES}_{i\vert\mathcal{J}_\sd}^{\tau_1\vert \tau_2}\equiv{\rm CoES}_{\tau_1}\left(\sY_i\mid\bY_{\mathcal{J}_\sd}=\widehat{\bpsi}_{\by_{\mathcal{J}_\sd}}\left(\hat{\by}_{\mathcal{J}_\sn}^{\tau_2}\right),\bY_{\mathcal{J}_\sn}=\widehat{\bpsi}_{\by_{\mathcal{J}_\sn}}\left(\hat{\by}_{\mathcal{J}_\sn}^{0.5}\right)\right)\nonumber
\end{equation}
is the Expected Shortfall of institution $i\in\mathcal{S}$, conditional on the set of institutions $\mathcal{J}_\sd$ being at their individual ${\rm ES}_{\tau_2}$-level $\widehat{\bpsi}_{\by_{\mathcal{J}_\sd}}\left(\hat{\by}_{\mathcal{J}_\sd}^{\tau_2}\right)=\left(\hat{\psi}_{\sy_{j_1}}\left(\hat{\sy}_{j_1}^{\tau_2}\right),\hat{\psi}_{\sy_{j_2}}\left(\hat{\sy}_{j_2}^{\tau_2}\right),\dots,\hat{\psi}_{\sy_{j_d}}\left(\hat{\sy}_{j_\sd}^{\tau_2}\right)\right)$ and the set of institutions $\mathcal{J}_\sn$ being at their individual ${\rm ES}_{0.5}$--level $\widehat{\bpsi}_{\by_{\mathcal{J}_\sn}}\left(\hat{\by}_{\mathcal{J}_\sn}^{\tau_2}\right)=\left(\hat{\psi}_{\sy_{j_{\sd+1}}}\left(\hat{\sy}_{j_1}^{\tau_2}\right),\hat{\psi}_{\sy_{j_2}}\left(\hat{\sy}_{j_2}^{\tau_2}\right),\dots,\hat{\psi}_{\sy_{j_{\sp-1}}}\left(\hat{\sy}_{j_{\sp-1}}^{\tau_2}\right)\right)$, with $\hat{\psi}_{\sy_{j}}\left(\hat{\sy}_{j}^{\tau}\right)\equiv{\rm ES}_{\tau}\left(\sY_j\right)$, $\forall j=1,2,\dots,\sd$, and can be defined in the following way
\begin{eqnarray}
{\rm MCoES}_{i\vert\mathcal{J}_\sd}^{\tau_1\vert \tau_2}\equiv\xp\left(\sY_i\mid\sY_i\leq\hat{\sy}_i^{\tau_1},\bY_{\mathcal{J}_\sd}=\widehat{\bpsi}_{\by_{\mathcal{J}_\sd}}\left(\hat{\by}_{\mathcal{J}_\sd}^{\tau_2}\right),\bY_{\mathcal{J}_\sn}=\widehat{\bpsi}_{\by_{\mathcal{J}_\sn}}\left(\hat{\by}_{\mathcal{J}_\sn}^{0.5}\right)\right).
\end{eqnarray}
\end{definition}
\noindent In the MS framework considered here the CoES reduces to the following weighted average
\begin{eqnarray}
{\rm MCoES}_{i\vert\mathcal{J}_\sd}^{\tau_1\vert \tau_2}
&=&\frac{1}{\tau_1}\sum_{l=1}^L\tilde{\eta}_l\left(\widehat{\bpsi}_{\by_{\mathcal{J}_\sd}}\left(\hat{\by}_{\mathcal{J}_\sd}^{\tau_2}\right),\widehat{\bpsi}_{\by_{\mathcal{J}_\sn}}\left(\hat{\by}_{\mathcal{J}_\sn}^{0.5}\right)\right)\nonumber\\
&&\qquad\qquad\qquad
\times\widehat{\bpsi}_{\sy_i}^{l}\left(\hat{\sy}_i^{\tau_1}\mid
\widehat{\bpsi}_{\by_{\mathcal{J}_\sd}}\left(\hat{\by}_{\mathcal{J}_\sd}^{\tau_2}\right),\widehat{\bpsi}_{\by_{\mathcal{J}_\sn}}\left(\hat{\by}_{\mathcal{J}_\sn}^{0.5}\right)\right),
\end{eqnarray}
where 
\begin{itemize}
\item[\textit{(i)}] $\widehat{\bpsi}_{\sy_i}^{l}\left(\hat{\sy}_i^{\tau_1}\mid
\widehat{\bpsi}_{\by_{\mathcal{J}_\sd}}\left(\hat{\by}_{\mathcal{J}_\sd}^{\tau_2}\right),\widehat{\bpsi}_{\by_{\mathcal{J}_\sn}}\left(\hat{\by}_{\mathcal{J}_\sn}^{0.5}\right)\right)$, for $l=1,2,\dots,L$ is the $l$--th component specific ES of the predictive distribution of $\sY_i$ conditional on the set of institutions $\mathcal{J}_\sd$ being at their individual ${\rm ES}_{\tau_2}$--level, $\bY_{\mathcal{J}_\sd}=\widehat{\bpsi}_{\by_{\mathcal{J}_\sd}}\left(\hat{\by}_{\mathcal{J}_\sd}^{\tau_2}\right)$ and the set of institutions $\mathcal{J}_\sn$ being at their individual ${\rm ES}_{0.5}$--level, $\bY_{\mathcal{J}_\sn}=\widehat{\bpsi}_{\by_{\mathcal{J}_\sn}}\left(\hat{\by}_{\mathcal{J}_\sn}^{0.5}\right)$.
\item[\textit{(ii)}] $\tilde{\eta}_l\left(\widehat{\bpsi}_{\by_{\mathcal{J}_\sd}}\left(\hat{\by}_{\mathcal{J}_\sd}^{\tau_2}\right),\widehat{\bpsi}_{\by_{\mathcal{J}_\sn}}\left(\hat{\by}_{\mathcal{J}_\sn}^{0.5}\right)\right)$ is the weight associated to the $l$--th component ES whose analytical expression can be found by plugging $\widehat{\bpsi}_{\by_{\mathcal{J}_\sd}}\left(\hat{\by}_{\mathcal{J}_\sd}^{\tau_2}\right)$ and $\widehat{\bpsi}_{\by_{\mathcal{J}_\sn}}\left(\hat{\by}_{\mathcal{J}_\sn}^{0.5}\right)$ into equation (\ref{eq:multivGaussMix_Cond_weights}) and (\ref{eq:multivStudentMix_Cond_weights}) for the multivariate Gaussian and and multivariate t MSM respectively.
\end{itemize}
%
%
\section{Individual contributions to overall risk}
\label{sec:systemic_risk_contribution}
%
When dealing with overall risk measures, it is important to quantify the marginal contribution of individual institutions to the overall risk. In their seminal paper, Adrian and Brunnermeier \cite{adrian_brunnermeier.2011} suggest that such marginal contribution can be evaluated by the $\Delta$CoVaR. Within their framework, where only two asset are considered, the $\Delta$CoVaR is defined as the difference between the CoVaR of institution $i$ conditional on institution $j$ being under distress and the CoVaR of the same institution $i$ when institution $j$ is at its median state. In this case the median state identifies the non--distress events of institution $j$. However, since during periods of financial instability several institutions may experience financial distress at the same time, following the same idea behind MCoVaR (MCoES), it is straightforward to generalise $\Delta$CoVaR ($\Delta$CoES) to Multiple--$\Delta$CoVaR (Multiple--$\Delta$CoES).
%
%
%
%
Hence, we define the ``Multiple--$\Delta$CoVaR'', $\Delta^\sM\text{CoVaR}_{i\vert\mathcal{J}_\sd}^{\tau_1\vert \tau_2}$ as follows:
\begin{eqnarray}
\Delta^\sM{\rm CoVaR}_{i\vert\mathcal{J}_\sd}^{\tau_1\vert \tau_2}&=&
{\rm CoVaR}_{\tau_1}\left(\sY_i\mid\bY_{\mathcal{J}_\sd}=\hat{\by}_{\mathcal{J}_\sd}^{\tau_2},\bY_{\mathcal{J}_\sn}=\hat{\by}_{\mathcal{J}_\sn}^{0.5}\right)-\nonumber\\
&&\qquad\qquad\qquad\qquad{\rm CoVaR}_{\tau_1}\left(\sY_i\mid\bY_{\mathcal{J}_\sd\cup\mathcal{J}_\sn}=\hat{\by}_{\mathcal{J}_\sd\cup\mathcal{J}_\sn}^{0.5}\right)
\label{eq:delta_m_covar}
\end{eqnarray}
$\forall i=1,2,\dots,\sp$. The $\Delta^\sM\text{CoVaR}_{i\vert\mathcal{J}_\sd}^{\tau_1\vert \tau_2}$ gives a more complete information on which combination of distressed institutions provides the largest contribution to the risk of the $i$-th institution and conveys a sharpened signal to the regulator.
%
%
It is worth noting that when $\mathcal{J}_\sd=\mathcal{S}\setminus\left\{i\right\}$ the proposed $\Delta^\sM{\rm CoVaR}$ risk measure quantifies the total risk contribution to the $i-$th institution, which is useful to assess the total amount of risk. Note also that when $\sd=1$, the $\Delta^{\sM}$CoVaR does not coincide with the Adrian and Brunnermeier \cite{adrian_brunnermeier.2011} $\Delta$CoVaR definition.\newline
\indent Following a similar idea, we can define the ``Multiple--$\Delta$CoES'', $\Delta^\sM{\rm CoES}_{i\vert\mathcal{J}_\sd}^{\tau_1\vert \tau_2}$, as follows
\begin{eqnarray}
\Delta^\sM{\rm CoES}_{i\vert\mathcal{J}_\sd}^{\tau_1\vert \tau_2}&=&
{\rm CoES}_{i\vert\mathcal{J}_\sd}^{\tau_1\vert \tau_2}\left(\sY_i\mid\bY_{\mathcal{J}_\sd}=\hat{\by}_{\mathcal{J}_\sd}^{\tau_2},\bY_{\mathcal{J}_\sn}=\hat{\by}_{\mathcal{J}_\sn}^{50\%}\right)-\nonumber\\
&&\qquad{\rm CoES}_{i\vert\mathcal{J}_\sd}^{\tau_1\vert \tau_2}\left(\sY_i\mid\sY_i\leq\hat{\psi}_{\sy_{i}}\left(\hat{\sy}_{i}^{\tau_1}\right),\bY_{\mathcal{J}_\sd\cup\mathcal{J}_\sn}=\hat{\by}_{\mathcal{J}_\sd\cup\mathcal{J}_\sn}^{50\%}\right).
\label{eq:delta_m_es}
\end{eqnarray}
\noindent In a multivariate environment, it is of fundamental importance to determine how the total risk shares among individual market participants. This can be accomplished by using the Shapley value theory, see e.g. Shapley \cite{shapley.1953} described in the next subsection which decomposes the total risk into individuals contributions.
\subsection{Shapley Value}
\label{subsec:shapley_value}
%
The Shapley value, initially formulated in a cooperative game theory approach, is used in this paper to efficiently allocate the overall risk among institutions belonging to the financial system. The Shapley value has been previously applied in a different context by Koylouglu and Stoker \cite{koylouglu_stoker.2002}, and as a measure to attributing systemic risk by Tarashev \textit{et al.} \cite{tarashev_etal.2010} and Cao \cite{cao.2013}.\newline
%
%
%
%
\indent The Shapley value methodology has been proposed to share utility or a cost among participant of a cooperative game where players can encourage cooperative behaviour and make coalitions. 
When the Shapley value is used as risk distributor the total gain of the coalition coincides with the overall risk generated by the financial system.\newline
\indent In a cooperative game, $\vartheta_i\left(\mathcal{H}\right)$, $i=1,2,\dots,\sp$, denotes the loss function of individual $i$, generated by group $\mathcal{H}\subset\mathcal{S}\setminus\left\{i\right\}$, where $\mathcal{S}=\left\{1,2,\dots,\sp\right\}$ as defined in the previous section and $\sp$ is the total number of institutions. In our risk measurement framework, the loss function $\vartheta_i:\mathbb{R}^{2^\sp}\longrightarrow\mathbb{R}^{+}$ coincides with the Multiple--$\Delta$CoVaR or the Multiple--$\Delta$CoES and assigns to each of the $2^{\sp-1}$ possible groups of institutions $\mathcal{H}$ its marginal contribution to the overall risk in such a way that $\vartheta_i\left(\mathcal{S}\setminus\left\{i\right\}\right)$ is the total risk of institution $i\in\mathcal{S}$. In the cooperative game theory framework the function $\vartheta_i\left(\cdot\right)$ should be super--additive and such that $\vartheta_i\left(\varnothing\right)=0$, where $\varnothing$ is the null set. This means that the contribution of a union of disjoint coalitions is not less than the sum of the coalition's separate values and that the contribution of an \qmo empty\qmcsp coalition is zero.\newline
\indent The Shapley value is one of the possible ways to distribute the total risk of institution $i\in\mathcal{S}$, i.e. $\vartheta_i\left(\mathcal{S}\setminus\left\{i\right\}\right)$, among all the remaining institutions belonging to the financial system assuming that they all collaborate. In particular, the Shapley value of institution $j\in\mathcal{S}\setminus\left\{i\right\}$, denoted by ${\rm ShV}_i\left(j\right)$, determines the amount of institution $j$'s risk contribution on institution $i$ and satisfies the \qmo\textit{individual rationality condition}\qmc, i.e. ${\rm ShV}_i\left(j\right)\geq \vartheta_i\left(\left\{j\right\}\right)$, $\forall i,j\in\mathcal{S}$, with $j\neq i$, where $\vartheta_i\left(\left\{j\right\}\right)$ is the marginal risk contribution of institution $j$ if it does not cooperate, and the \qmo\textit{collective rationality condition}\qmc, i.e. $\sum_{j=1}^\sp{\rm ShV}_i\left(j\right)=\vartheta_i\left(\mathcal{S}\right)$, $\forall i\in\mathcal{S}$.
The Shapley values are obtained as
\begin{eqnarray}
{\rm ShV}_i\left(j\right)=\frac{1}{\vert\mathcal{S}\setminus\left\{i\right\}\vert!}\sum_{\mathcal{H}\subset\mathcal{S}\setminus\left\{i,j\right\}}
\vert\mathcal{H}\vert!\left(\vert\mathcal{S}\setminus\left\{i\right\}\vert-\vert\mathcal{H}\vert-1\right)!
\left[\vartheta_i\left(\mathcal{H}\cup\left\{j\right\}\right)-\vartheta_i\left(\mathcal{H}\right)\right]
\end{eqnarray}
for $i,j=1,2,\dots,\sp$ with $j\neq i$, where the sum extends over all the subsets $\mathcal{H}$ of $\mathcal{S}\setminus\left\{i\right\}$ not containing institution $j$. The Shapley values are the shares of each institution from the value of the loss function when all players cooperate and 
%
%
possesses the following desirable properties.
\begin{itemize}
\item[1.] \textit{Efficiency}: $\sum_{j=1}^\sp{\rm ShV}_i\left(j\right)=\vartheta_i\left(\mathcal{S}\setminus\left\{i\right\}\right)$, $\forall i\in\mathcal{S}$. The efficiency axiom states that the total risk of institution $i$ is distributed among all the remaining market participants with no loss and no gains. The overwhelming importance of this property can be clearly understood when considering macro prudential regulation. Additive risk measures imply that supervisors will not penalise the economy for no reasons.
%
%
\item[2.] \textit{Symmetry}: if $k\neq i,j$ such that $\vartheta_i\left(\mathcal{H}\cup\left\{j\right\}\right)=\vartheta_i\left(\mathcal{H}\cup\left\{k\right\}\right)$, $\forall\mathcal{H}$ such that $i,j,k\notin\mathcal{H}$, then ${\rm ShV}_i\left(j\right)={\rm ShV}_i\left(k\right)$. The symmetry axioms simply states that the Shapley value is permutation invariant and induces a fairness  property of the resulting risk distributor. This means that if the marginal risk contribution of any two institutions and for any subset $\mathcal{H}$ is the same, then their Shapley values should be the same.
\item[3.] \textit{Dummy axiom}: if $\vartheta_i\left(\mathcal{H}\cup\left\{j\right\}\right)=\vartheta_i\left(\left\{j\right\}\right)$, $\forall j\in \mathcal{H}$ and $\mathcal{H}\not\supset i$, then ${\rm ShV}_i\left(j\right)=\vartheta_i\left(\left\{j\right\}\right)$. This property means that if the  risk of institution $j$ is independent of any other institutions' risk conditionally on institution $i$, then the risk share of $j$ should be exactly equal to its risk alone. This is the case where standard CoVaR of Adrian and Brunnermeier \cite{adrian_brunnermeier.2011} coincides with the Shapley value methodology. As argued by Cao \cite{cao.2013}, since in general the risk measure of institution $j$ is not conditionally orthogonal of any other institutions' risk, this implies that the standard CoVaR approach and the Shapley value marginal contribution does not delivers the same risk ordering.
\item[4.] \textit{Linearity (or additivity)}: If $j$ and $k$, with $j,k\in\mathcal{H}$ and $j\neq k$ are two different institutions described by the functions $\vartheta_i\left(j\right)$ and $\vartheta_i\left(k\right)$, with $\vartheta_i\left(j\right)\neq\vartheta_i\left(k\right)$, such that their linear combination delivers a new function $\tilde{\vartheta}_i=w_j\vartheta_i\left(j\right)+w_k\vartheta_i\left(k\right)$, with $w_l>0, \forall l=\left\{j,k\right\}$, then the distributed risks should equal the weighted average of individual risk contribution as evaluated by their respective Shapley values, i.e. ${\rm ShV}_i\left(\tilde{\theta}_i\right)=w_j{\rm ShV}_i\left(j\right)+w_k{\rm ShV}_i\left(k\right)$.
%
%
\item[5.] \textit{Zero player}: a player is null if none of the subsets $\mathcal{H}$ contains $j$. A null player receives zero risk contribution. 
\end{itemize}
%
%
%
%
\section{Empirical Analysis}
\label{sec:empirical_analysis}
%
We apply the econometric framework and the methodology described in previous sections to examine the systemic risk in the US banking system.
%
\subsection{Data}
\label{sec:data}
\begin{figure}[!t]
\begin{center}
\captionsetup{font={small}, labelfont=sc}
\includegraphics[width=1.00\linewidth]{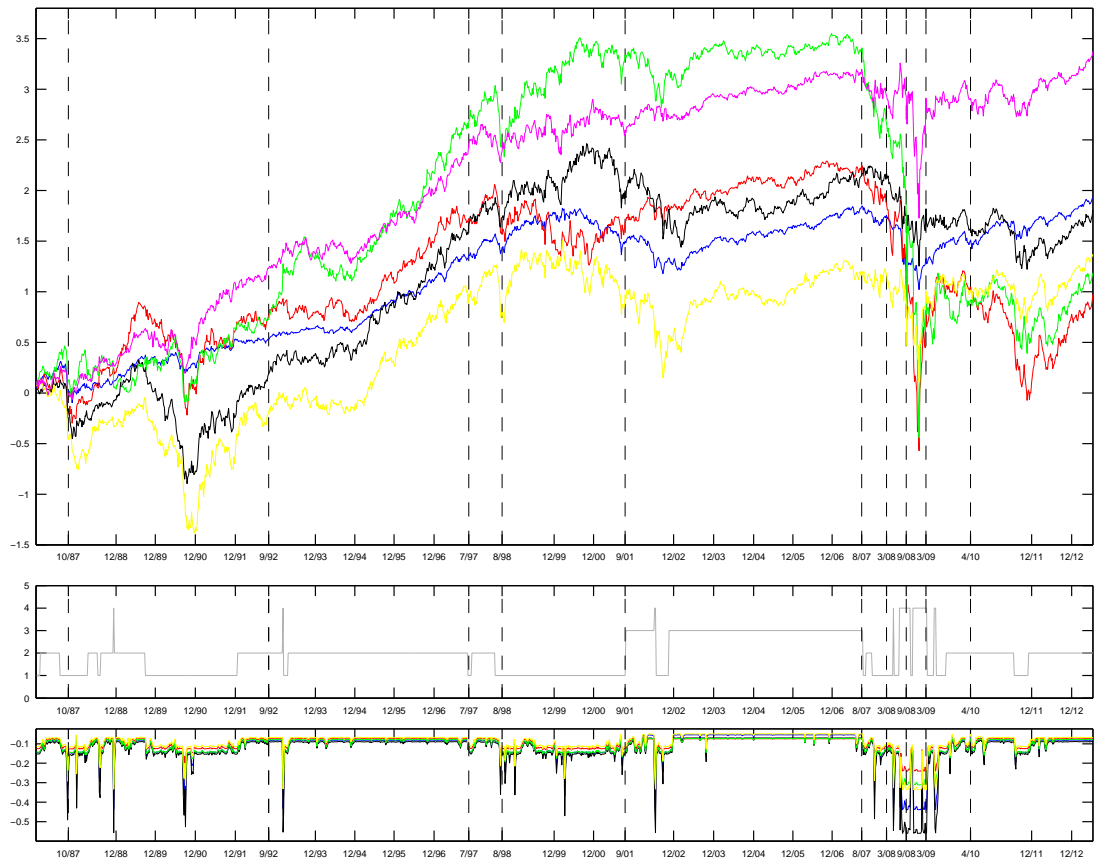}
\caption{\small{\textit{(Top panel)}: cumulative returns of the BAC (red line), BK (dark line), C (green line), JPM (yellow line), WFC (magenta) and S\&P500 index (dark line), since January 1st, 1987 till the end of the period. \textit{(Middle panel)}: smoothed hidden states. \textit{(Bottom panel)}: Expected Shortfall at the 5\% level.
Vertical dotted lines represent major financial downturns: the ``Black Monday'', (October 19, 1987), the ``Black Wednesday'' (September 16, 1992), the Asian crisis (July, 1997), the Russian crisis (August, 1998), the September 11, 2001 shock, the Bear Stearns hedge funds collapse (August 5, 2007), the Bear Stearns acquisition by JP Morgan Chase, (March 16, 2008), the Lehman's failure (September 15, 2008), the peak of the onset of the recent global financial crisis (March 9, 2009) and the European sovereign-debt crisis of April 2010 (April 23, 2010, Greek crisis).}}
\label{fig:International_cum_returns}
\end{center}
\end{figure}

%
We consider a panel of US Banks belonging to the Standard and Poor's Composite Index, S\&P500. The basket consists on weekly returns of five among the major US Banks by capitalization, covering the period from January 2nd, 1987 to June 28th, 2013: BAC (Bank of America Corp), BK (The Bank of New York Mellow Corp.), C (Citigroup Inc.), JPM (JPMorgan Chase \& Co.) and WFC (Wells Fargo). All the time series are from the Bloomberg Database. Descriptive statistics for the data are provided in Table \ref{tab:USBanks_data_summary_stat}. In line with stylised facts of financial time series, the returns are positively skewed (except for Citigroup (C), which is hugely negatively skewed) and leptokurtic, indicating that they are not normally distributed. In addition, the Jarque--Bera (JB) statistic confirms the departure from normality for all return series at the 1\% level of significance. Moreover, the presence of large volatility clusters followed by periods of low volatility is documented by the data. This facts are coherent with the presence of different regimes of \qmo bull\qmcsp and \qmo bear\qmcsp market conditions.\newline
\indent The goal is to analyse how the systemic risk spreads among the different institutions by inspecting the time evolution of the risk measures introduced in previous sections.
%
\begin{table}[!t]
\captionsetup{font={small}, labelfont=sc}
%
\begin{small}
\resizebox{\columnwidth}{!}{%
\centering
 \smallskip
  \begin{tabular}{lcccccccc}\\
  \hline\hline
   Name & Min & Max & Mean$\times10^3$ & Std. Dev. & Skewness & Kurtosis & 1\% Str. Lev. & JB\\
    \hline
BAC& -0.593  & 0.607  & 0.697  & 0.059  & -0.277  & 28.671  & -0.151  & 38431.115  \\
BK& -0.246  & 0.260  & 1.291  & 0.046  & -0.061  & 6.121  & -0.130  & 568.713  \\
C& -0.926  & 0.788  & 0.847  & 0.069  & -1.422  & 51.510  & -0.155  & 137644.599  \\
JPM& -0.417  & 0.399  & 0.980  & 0.054  & -0.175  & 11.499  & -0.130  & 4217.499  \\
WFC& -0.368  & 0.482  & 2.411  & 0.047  & 0.273  & 20.971  & -0.123  & 18842.051  \\
S\&P500& -0.201  & 0.114  & 1.377  & 0.024  & -0.860  & 9.600  & -0.070  & 2711.287  \\
      \hline\hline
\end{tabular}}
\caption{Summary statistics of five US banks in the panel and the SP\&500 index, for the period form January, 2nd 1987 till June, 28th 2013. The eight column, denoted by ``1\% Str. Lev.'' is the 1\% empirical quantile of the returns distribution, while the last column, denoted by ``JB'' is the value of the Jarque-Ber\'a test-statistics.}
\label{tab:USBanks_data_summary_stat}
\end{small}
%
%
\end{table}
%
\noindent We would examine whether stock market co-movements have changed over time, with a focus on the crisis periods. On the one hand, we expect that the global nature of the financial crisis might imply that the co--movements become stronger, with an increase in the long--run risks. On the other hand, given the heterogeneous composition of the considered panel, where institutions strongly differ by market capitalisation and other individual characteristics, (some observable like the business core, some other not observable like debt composition or the level of market linkage)
we would expect that different banks were hit rather unequally by the 2007--2008 global financial crisis. For example, by looking at the kurtosis index in Table \ref{tab:USBanks_data_summary_stat} we observe that Citigroup (C) has much larger fat tails than other banks in the panel, and may be more affected by the financial crisis than other banks do.\newline
\indent The top panel in Figure \ref{fig:International_cum_returns} shows the time series of cumulative returns for all the considered assets, from January 2nd, 1987 till the end of the sample. Vertical dotted lines refer to the following events: the ``Black Monday'' (October 19, 1987), the ``Black Wednesday'' (September 16, 1992), the Asian crisis (July, 1997), the Russian crisis (August, 1998), the September 11--2001 shock, the onset of the mortgage subprime crisis identified by the Bear Stearns hedge funds collapse (August 5, 2007), the Bear Stearns acquisition by JP Morgan Chase, (March 16, 2008) and the Lehman's failure (September 15, 2008), the peak of the onset of the recent global financial crisis (March 9, 2009) and the European sovereign--debt crisis of April 2010 (April 23, 2010, Greek crisis). The figure gives insights about the effect of crisis periods in each institutions and the overall market represented by the S\&P500 composite index (dark dotted line). After the 2001 Twin Towers attack till the middle of 2007, the US financial system experienced a long period of small perturbations and stability ended shortly after the collapse of two Bear Stearns hedge funds in early August 2007. Starting from August 2007, the financial market experiences a huge fall, the subprime mortgage crisis that led to a financial crisis and subsequent recession that began in 2008. Several major financial institutions collapsed in September 2008, with significant disruption in the flow of credit to businesses and consumers and the onset of a severe global recession. The system hit the bottom in March 2009, and then started a slow recovery which culminated just before the European sovereign--debt crisis of April 2010. It is interesting to note that, since the beginning of the 2007 global crisis all the considered institutions, as well as the market index, experienced huge capital losses with Citigroup (C) (green line) being the most affected by the crisis. Moreover, Citigroup (C), JP Morgan (JPM) (yellow line) and Bank of America Corp (BAC) (red line) become more correlated after European sovereign--debt crisis, displaying similar trends.
%
%

%
\subsection{Full sample estimation results}
\label{sec:estimation_results}
%
To account for all observed data features, we estimate multivariate Gaussian and Student--t Markov Switching models over the entire sample period. 
A fundamental problem in fitting MS models is the choice of the number of latent states. 
%
%
In the literature on latent variables, the most used model selection tools are the Akaike Information Criterion (AIC) and the Bayesian Information Criterion (BIC). 
These two indexes involve penalisation terms depending on the number of non--redundant parameters (see e.g. Ryd\'en \cite{ryden.2008}).\newline
%
%
\indent In Table \ref{tab:DataSet_Model_Selection} we report the maximum log-likelihood, AIC, and BIC for Gaussian and Student--t MSMs when the number of hidden states varies from 2 to 5.
%
%
For each value of $L$, we use 20 random starting points to initialise model parameters, and we report the results corresponding to the best solution in terms of log--likelihood. We stop the algorithms when the increase in the log--likelihood is less than $10^{-5}$; for each of the two models, whenever possible, we use the same starting points.\newline
%
\begin{table}[!t]
\captionsetup{font={small}, labelfont=sc}
%
\begin{small}
\centering
 \smallskip
  \begin{tabular}{cccc}\\
  \hline\hline
  \multicolumn{4}{c}{\textbf{Multivariate Gaussian--MSM}}\\
\cmidrule(lr){2-4}
   L & log-likelihood & AIC & BIC \\
 \cmidrule(lr){1-1} \cmidrule(lr){2-2} \cmidrule(lr){3-3} \cmidrule(lr){4-4}
 2  & 132909.573  & -265633.146  & -265048.593  \\
3  & 133511.106  & -266736.211  & -265837.383  \\
4  &134106.228  & -267714.455  & -265820.188  \\
5  &134250.024   & \textbf{-268101.829}  &\textbf{-266149.362}   \\
6  & 134413.915   & -267890.048  & -265972.966   \\
\hline
\multicolumn{4}{c}{\textbf{Multivariate t--MSM}}\\
\cmidrule(lr){2-4}
   L & log-likelihood & AIC & BIC  \\
 \cmidrule(lr){1-1} \cmidrule(lr){2-2} \cmidrule(lr){3-3} \cmidrule(lr){4-4} 
2  & 137686.512  & -275283.023  & -274975.669    \\
3  & 138085.711  & -276029.421  & -275544.484   \\
4  & 138402.144  & \textbf{-277714.991}  & \textbf{-276137.237}   \\
5  & 139088.495  & -276546.288  & -275665.205    \\
6  & 138672.683  & -277023.366  & -275923.720    \\
      \hline\hline
\end{tabular}
\caption{Log-likelihood, AIC and BIC values for the multivariate Gaussian (MVN) and Student-t (MVT) Hidden Markov models fitted to the panel of US banks and the S\&P500 index. Bold faces indicates the selected model.}
\label{tab:DataSet_Model_Selection}
\end{small}
%
%
\end{table}
%
\noindent Table \ref{tab:DataSet_Model_Selection} highlights that the Student--t MSMs account for the asymmetry and kurtosis displayed by the observed data in a more parsimonious way as compared to the Gaussian alternatives. In fact, the selected model is the Student--t with $L=4$.
%
%
%
\noindent Table \ref{tab:DataSet_International_Estimates_1} and \ref{tab:DataSet_International_Estimates_2} summarise parameter estimates for the chosen model, which all display typical features common to return series. 
As expected, hidden regimes are identified by volatility. Despite of the common findings state--specific return means differ significantly across latent states, favouring the rejection of the null hypothesis that the conditional means are equal.
%
%
Means and variances differences across state exacerbate the visited state persistence observed in our time series data. This can be evinced also by inspecting the middle panel of Figure \ref{fig:International_cum_returns} depicting the Viterbi estimates of the regimes: the selected MSM is able to recover the underlying structure, capturing periods of crisis as well as stable phases. Indeed, according to state--specific return means, we identify two positive and two negative regimes. 
Furthermore, during a financial crisis (identified by State 1 and State 4), stock returns experiences high negative average mean returns, and variances are pretty large. Vice versa, during more stable phases (as those identified by State 2 and State 3), instead, stock returns fluctuate around a  positive mean, and variances are relatively low.
%
%
As a consequence, the selected multivariate MS Student--t model is able to distinguish and cluster time periods corresponding to different risk--returns profiles.
%
%
This evidence is confirmed by observing the small estimated values for the $\boldsymbol{\nu}$ parameters detecting the presence of fat tails. Moreover, the estimated transition probability matrix $\boldsymbol{Q}$, which explains the evolution over time of state--switching, well captures the low rate of regime switching. In particular, off--diagonal probabilities are generally low. An interesting aspect is related to crisis periods: bursts in the crisis (as those identified in State 4) can be likely followed by another crisis period with less marked losses. Thus, coming out suddenly from a crisis period is unlikely. As known, the persistence of regimes plays an important role in generating volatility clustering, for which periods of high volatility are followed by high volatility, and periods of low volatility are followed by low volatility.\newline
%
%
\indent Due to the multivariate approach considered here, we are able to identify different spillover effects among stocks, measured by the state--specific correlations. As expected, correlations are higher during crisis period than during more stable phases. This evidence has several important consequences for the tail risk interdependence analysed in the next subsection.
%
%
%
Nevertheless, even if parameter estimates can support the policy decision making process they do not provide enough information to evaluate extreme tail interdependence between stocks, and risk measures should be constructed on the basis of the obtained parameter estimates.
%
\begin{table}[!t]
\captionsetup{font={small}, labelfont=sc}
%
\begin{small}
\centering
 \smallskip
  \begin{tabular}{ccccccc}\\
  \hline\hline
  $\boldsymbol{\mu}\times 10^3$&SPX& BAC & BK & C & JPM & WFC\\
\hline
State 1& 0.6888 & -3.4629 & -1.8248 & -1.3567 & -3.4742 & 0.6349 \\
State 2& 3.4713 & 4.3184 & 4.9317 & 5.2207 & 4.9884 & 3.9972 \\
State 3& 1.6648 & 1.8952 & 0.5980 & 0.0617 & 1.1678 & 0.8968 \\
State 4& -9.3863 & -13.9701 & -12.1881 & -23.4741 & -9.2301 & 2.5614 \\
\hline
$\boldsymbol{\Lambda}\times 10^3$ &SPX& BAC & BK & C & JPM & WFC\\
\hline
State 1& 0.6730 & 3.0970 & 2.6547 & 3.3052 & 3.3422 & 2.0985 \\
State 2& 0.2341 & 1.1694 & 0.9673 & 1.4067 & 1.1136 & 0.8347 \\
State 3& 0.2285 & 0.3153 & 0.7392 & 0.5201 & 0.6806 & 0.2819 \\
State 4& 2.8634 & 37.8412 & 10.9080 & 60.3287 & 19.0749 & 23.9363 \\
\hline
$\boldsymbol{\Omega}_1$ &SPX& BAC & BK & C & JPM & WFC\\
\hline
SPX & 1.0000 & & & & & \\
BAC & 0.6127 & 1.0000 & & & & \\
BK & 0.5757 & 0.5842 & 1.0000 & & & \\
C & 0.7242 & 0.6345 & 0.5504 & 1.0000 & & \\
JPM & 0.6217 & 0.6739 & 0.6226 & 0.6385 & 1.0000 \\
WFC & 0.5818 & 0.6528 & 0.5595 & 0.5552 & 0.5690 & 1.0000 \\
\hline
$\boldsymbol{\Omega}_2$ & SPX& BAC & BK & C & JPM & WFC\\
\hline
SPX & 1.0000 & & & & & \\
BAC & 0.5943 & 1.0000 & & & & \\
BK & 0.6213 & 0.6104 & 1.0000 & & & \\
C & 0.6459 & 0.5798 & 0.5183 & 1.0000 & & \\
JPM & 0.6050 & 0.6334 & 0.6032 & 0.5545 & 1.0000 \\
WFC & 0.6670 & 0.5853 & 0.6008 & 0.5348 & 0.6035 & 1.0000 \\
\hline
$\boldsymbol{\Omega}_3$ & SPX& BAC & BK & C & JPM & WFC\\
\hline
SPX & 1.0000 & & & & & \\
BAC & 0.6247 & 1.0000 & & & & \\
BK & 0.6979 & 0.4915 & 1.0000 & & & \\
C & 0.7316 & 0.6961 & 0.5790 & 1.0000 & & \\
JPM & 0.7751 & 0.6537 & 0.6762 & 0.7447 & 1.0000 \\
WFC & 0.6172 & 0.7156 & 0.5087 & 0.6695 & 0.6359 & 1.0000 \\
\hline
$\boldsymbol{\Omega}_4$ & SPX& BAC & BK & C & JPM & WFC\\
\hline
SPX & 1.0000 & & & & & \\
BAC & 0.7366 & 1.0000 & & & & \\
BK & 0.7447 & 0.7215 & 1.0000 & & & \\
C & 0.6688 & 0.8040 & 0.6177 & 1.0000 & & \\
JPM & 0.7198 & 0.7728 & 0.7208 & 0.7613 & 1.0000 \\
WFC & 0.6798 & 0.8731 & 0.7579 & 0.7460 & 0.8287 & 1.0000 \\
\hline\hline
\end{tabular}\\
\caption{ML parameter estimates of the selected Multivariate t--MSM with four components where $\boldsymbol{\mu}$ are locations while the diagonal matrix $\bLambda$ and the full matrix $\bOmega$ are such that $\bSigma=\bLambda\bOmega\bLambda$.}
\label{tab:DataSet_International_Estimates_1}
\end{small}
%
%
\end{table}
%
%
%
\begin{table}[!t]
\captionsetup{font={small}, labelfont=sc}
%
\begin{small}
\centering
 \smallskip
  \begin{tabular}{cccccc}\\
  \hline\hline
 \multicolumn{1}{c}{\multirow{2}{*}{$\boldsymbol{\nu}$}}& State 1 & State 2 &State 3 & State 4\\
 \cmidrule(lr){2-2} \cmidrule(lr){3-3} \cmidrule(lr){4-4} \cmidrule(lr){5-5}
 & 11.8208 & 13.1583 & 6.5634 & 13.8517 \\
\hline
\multicolumn{1}{c}{\multirow{2}{*}{$\boldsymbol{\delta}$}}& State 1 & State 2 &State 3 & State 4\\ \cmidrule(lr){2-2} \cmidrule(lr){3-3} \cmidrule(lr){4-4} \cmidrule(lr){5-5}
 & 1.0000 & 0.0000 & 0.0000 & 0.0000 \\
\hline
$\mathbf{Q}$& State 1 & State 2 &State 3 & State 4\\
\cmidrule(lr){1-1} \cmidrule(lr){2-2} \cmidrule(lr){3-3} \cmidrule(lr){4-4} \cmidrule(lr){5-5}
State 1& 0.9485 & 0.0302 & 0.0088 & 0.0125 \\
State 2& 0.0185 & 0.9785 & 0.0000 & 0.0030 \\
State 3& 0.0090 & 0.0000 & 0.9865 & 0.0045 \\
State 4& 0.1721 & 0.0221 & 0.0000 & 0.8057 \\
\hline\hline
\end{tabular}\\
\caption{ML estimates of the initial probability $\boldsymbol{\delta}$ and transition probability matrix $\mathbf{Q}$ of the Markov chain for the selected MVT hidden Markov models.}
\label{tab:DataSet_International_Estimates_2}
\end{small}
%
%
\end{table}
%

%
%
\subsection{Marginal contribution to systemic risk}
\label{sec:systemic_risk}
\begin{figure}[!t]
\begin{center}
\captionsetup{font={small}, labelfont=sc}
\includegraphics[width=1.0\linewidth]{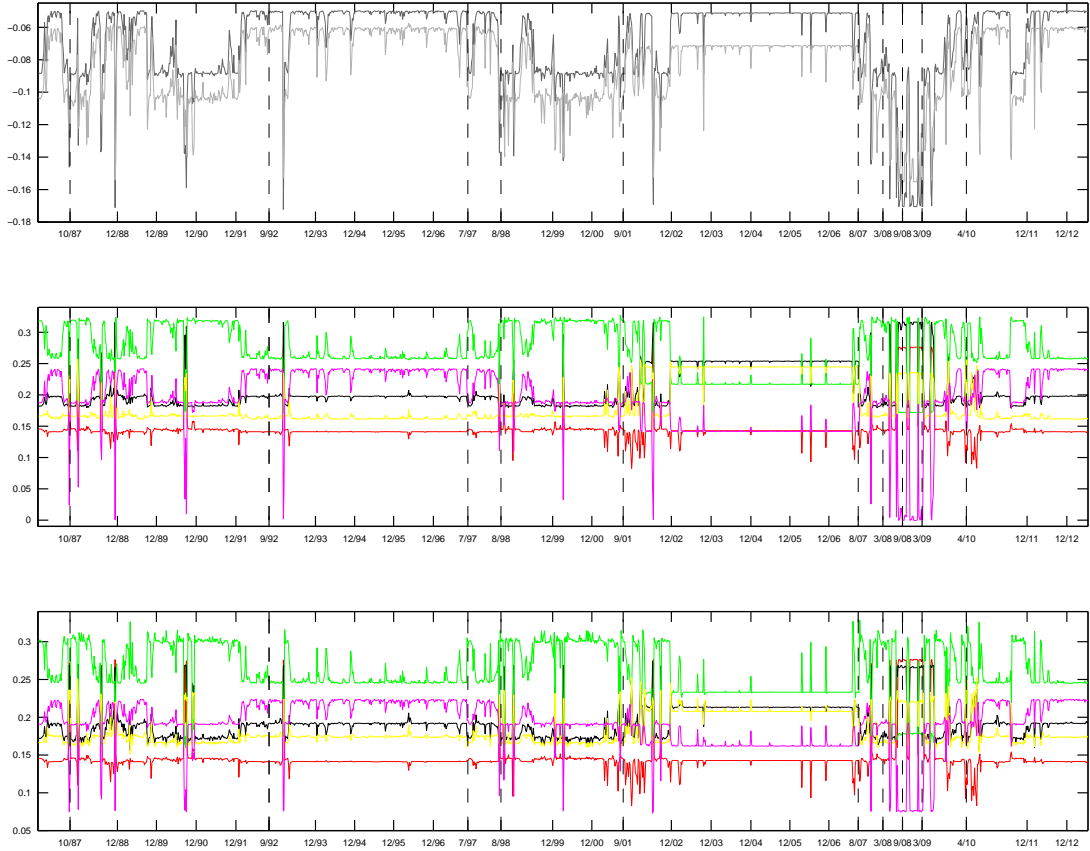}
\caption{\small{\textit{(Top panel):} total systemic risk measured by $\Delta\text{CoVaR}_{k\vert S}^{\tau_1\vert \tau_1}$ (gray line) and by $\Delta\text{CoES}_{k\vert S}^{\tau_1\vert \tau_1}$ (light gray line), where $k$ denotes the S\&P500 index and $S$ denotes the set of indexes of the remaining assets. \textit{(Middle panel):} marginal contribution to systemic risk evaluated by means of the Shapley value methodology ${\rm ShV_i}$ based on $\Delta$CoVaR, for all the banks in the panel. \textit{(Bottom panel):} marginal contribution to systemic risk evaluated by means of the Shapley value methodology ${\rm ShV_i}$ based on $\Delta$CoES, for all the banks in the panel.  BAC (red line), BK (dark line), C (green line), JPM (yellow line) and WFC (magenta line).\newline
Vertical dotted lines represent major financial downturns: the ``Black Monday'', (October 19, 1987), the ``Black Wednesday'' (September 16, 1992), the Asian crisis (July, 1997), the Russian crisis (August, 1998), the September 11, 2001 shock, the Bear Stearns hedge funds collapse (August 5, 2007), the Bear Stearns acquisition by JP Morgan Chase, (March 16, 2008), the Lehman's failure (September 15, 2008), the peak of the onset of the recent global financial crisis (March 9, 2009) and the European sovereign-debt crisis of April 2010 (April 23, 2010, Greek crisis).}}
\label{fig:USBanks_Shapley_value}
\end{center}
\end{figure}

%
One of the most important questions a systemic risk measure should answers is the identification of institutions that are systematically more important or contribute more to the vulnerability of the overall system. Using the methodology described in previous sections, we are able to measure individual institutions' systemic importance at each point in time as well as their variation over time, especially during the financial crisis. Moreover, such risk measures provide important monitoring tools for the market--based macro--prudential or financial stability regulation.\newline
\indent For the basket of assets considered, we calculate the marginal contributions of each institution to the overall systemic risk by means of the Shapley value. The top panel of Figure \ref{fig:USBanks_Shapley_value} plots the overall systemic risk based on $\Delta^\sM$CoVaR and $\Delta^\sM$CoES at $\tau_1=\tau_2=0.05$. The total systemic risk is at its minimum level before the Bear Stearns hedge funds collapse (August 5, 2007) and then increases significantly during year 2007 till the middle of 2008, when Lehaman and Brothers failed (September 2009). Subsequently, the system experienced a long period (between September 2008 and March 2009) of financial instability and high volatility when the overall risk is at its highest level. Then the total systemic risk decreased suddenly, reaching the pre--Bear Sterns collapse level in the middle of 2009. This probably has been the major consequence of the US Supervisory Capital Assessment Program (SCAP) conducted by the Federal Reserve System to determine if the largest US financial organisations had sufficient capital buffers to withstand the recession and the financial market turmoil whose results were released on May 7, 2009. The market has calmed down till the first round of European sovereign debt crisis in May 2010, after the Greece receiving the aid with 14,5 billions euros, as documented by the decrease of the total systemic risk. Almost a year later, (June 13, 2011), Standard \& Poor's has downgraded Greek debt from B to CCC, and the total systemic risk raised sharply reaching the higher peak after 2007--2009 financial crisis in summer 2011.\newline
\indent Concerning the total systemic risk exposure during the recent global financial crisis of 2007--2008, we observe that it is about three time as much larger as it was during the previous period (2002--2007) of financial stability and about two time larger than it was during the 1987 financial crisis, except for the black Monday week, when we observe pretty the same level as in 2008.\newline
\indent The medium and bottom panels of Figure \ref{fig:USBanks_Shapley_value} plot the individual marginal contribution to the systemic risk calculated by means of the Shapley value ${\rm ShV}_i$ based on $\Delta^{\sM}{\rm CoVaR}_{i\vert\mathcal{J}_\sd}$ and $\Delta^{\sM}{\rm CoES}_{i\vert\mathcal{J}_\sd}$ respectively. Here, the index $i$ denotes each individual institutions' risk contribution calculated with respect to the market index (S\&P500). These two panels track the systemic risk importance of each financial institution in percentage points. The level of systemic importance changes over time and in particular during period of financial instability. Looking for example at the bottom panel, displaying the ${\rm ShV}_i$ based on the $\Delta^{\sM}{\rm CoES}_{i\vert\mathcal{J}_\sd}$, we observe that Citigroup (C), (green line) is the bank that have the most systemic importance weights during the whole period, whereas the systemic weights of all the remaining institutions in the panel change quite a lot during period of financial turbulence as compared to their level observed during periods of financial stability. In particular, before the 2007--2008 financial crisis Bank of America (BAC) (red line) and Wells Fargo (WFC) (magenta line), display the lowest systemic risk contribution, followed immediately by JP Morgan (JPM) (yellow line) and Bank of NY Mellow (BK) (dark line) having the larger risk contributions. During the 2007--2008 financial crisis we observe two distinct phenomena: the ordering of systemic importance and the level of individual systemic importance change compare to the pre-crisis values. For example Wells Fargo (WFC) becomes the less important during the crisis period; moreover some institutions, like for example Bank of America, BAC, increase its systemic contribution while others, like for example Wells Fargo (WFC) reduce it. Comparing the middle and the bottom panel of Figure \ref{fig:USBanks_Shapley_value} we observe some important differences concerning the ordering of systemic importance during periods of crisis as compared with the non--crisis periods. This means that the choice of the systemic risk measure effectively impacts the risk assessment process and suggests that the subadditivity property should be taken into consideration even for systemic risk measurement purposes. In fact, the documented discordance between the two risk measures can be probably ascribed to the known coherence deficiency of the VaR that rebounds on the CoVaR. Table \ref{tab:Shapley_SS} summarises the Shapley value ${\rm ShV}_i$ statistics based on $\Delta^{\sM}{\rm CoVaR}_{i\vert\mathcal{J}_\sd}$ and $\Delta^{\sM}{\rm CoES}_{i\vert\mathcal{J}_\sd}$ respectively. The provided statistics are conditional on the Markovian state identified by means of the Viterbi algorithm.

%
%

\begin{table}[!t]
\begin{small}
\begin{center}\tabcolsep=2.0mm
    \begin{tabular}{lcccc}
    \hline
    \hline
 & \multicolumn{4}{c}{\textit{Mean}}\\
\cmidrule(lr){1-1}\cmidrule(lr){2-5}   
 Bank & State 1 & State 2 & State 3 & State 4\\
 \cmidrule(lr){1-1}\cmidrule(lr){2-2}\cmidrule(lr){3-3}\cmidrule(lr){4-4}\cmidrule(lr){5-5}
BAC 	 	& 17.27\% & 16.45\% & 17.65\% & 25.14\% \\
BK 		& 18.02\% & 19.04\% & 21.10\% & 26.04\% \\
C 		& 28.86\% & 25.47\% & 24.04\% & 18.41\% \\
JPM 		& 17.22\% & 17.55\% & 20.58\% & 21.90\% \\
WFC 	& 18.63\% & 21.49\% & 16.64\% & 8.51\% \\
\hline
 & \multicolumn{4}{c}{\textit{Variance}}\\
\cmidrule(lr){1-1}\cmidrule(lr){2-5}   
 Bank & State 1 & State 2 & State 3 & State 4\\
 \cmidrule(lr){1-1}\cmidrule(lr){2-2}\cmidrule(lr){3-3}\cmidrule(lr){4-4}\cmidrule(lr){5-5}
BAC		& 1.80\% &  0.85\% &  1.89\% & 2.22\% \\
BK 		& 2.29\% &  0.87\% &  0.95\% & 2.21\% \\
C		& 3.41\% &  1.73\% &  1.99\% & 2.67\% \\
JPM		& 1.73\% &  0.97\% &  1.18\% & 1.11\% \\
WFC		& 2.32\% &  1.16\% &  1.03\% & 2.78\% \\
          \hline
         \hline
\end{tabular}

  \caption{\small{Summary statistics of the Shapley value.}}
  \label{tab:Shapley_SS}
    \end{center}
    \end{small}
\end{table}
%

%
\section{Conclusion}
\label{sec:conclusion}
%
In this paper, we develop a multivariate model--based approach to measure tail risk interdependences among institutions. 
We consider both Gaussian and Student--t Markov Switching models accounting for multiple underlying risk--return profiles. The proposed approach considers new tail interdependence risk measures $\Delta^\sM$CoVaR and $\Delta^\sM$CoES which naturally extend and improve the Adrian and Brunnermeier \cite{adrian_brunnermeier.2011} ideas of $\Delta$CoVaR and $\Delta$CoES to the multiple joint occurrence of extreme distress events.
%
Analytical expressions for those measures are evaluated on the predictive distributions in order to provide a forward--looking risk quantification. The Shapley value methodology is then applied to combine those multiple risk measures into an overall risk indicator that essentially distributes the risk among the market participants.
The idea behind this paper is to measure extreme risks characterised by a dynamic Markov Switching evolution being able, at the same time, to capture the main empirical evidence of stylised facts of financial returns. 
The developed methodology is applied to five major US banks belonging to the Standard and Poor's 500 index in order to assess individual institutions' marginal contribution to the systemic risk.
Comparing the Student--t and Gaussian results, we observe that the former assumption is preferred by standard information criteria and this supports the use of fat tailed distributions for financial data. Moreover, the choice of the Student--t distribution is also justified by theoretical reasons because it allows to model non linear dependence among tail events which is not possible under the multivariate Gaussian assumption.
Our empirical results suggest that the marginal contribution to the systemic risk of individual banks varies through time and particularly during periods of financial crisis it changes dramatically, both in order of importance and in levels. A decomposition analysis shows that the marginal contribution of individual banks to the systemic risk are determined mostly by bank size, consistent with the \qmo too-big-to-fail\qmcsp paradigm.
%
%
More importantly, we observe that merging a model being able to correctly identify different volatility regimes
%
with systemic risk measures that account for contemporaneous multiple distresses, allows us to well understand and predict the impact of financial market turmoils on individual and systemic risks.\newline
\indent 
Concluding, our analysis provides useful suggestions for the ongoing discussion on the imposition of capital requirements on systemically important institutions to prevent financial system disasters spillover effects. Concerning this aspect its main implication is that capital requirements should be based on forward looking risk measures being able to account for the evolving economic and financial conditions. 
The model can be extended to include exogenous information on individual institutions or by considering long--run forecasting horizons which implies the redefinition of some of the considered risk measures.
\newpage
\appendix
\section{Expectation--Maximization algorithm}
\label{sec:appendix_A}
%
\begin{description}
\item[E--step:] at iteration $\left(m+1\right)$, the E--step requires the computation of the so--called $\mathcal{Q}$--function, which calculates the conditional expectation of the complete--data log--likelihood given the observations and the current parameter estimates $\boldsymbol{\theta}^{\left(m\right)}$
\begin{eqnarray}
\mathcal{Q}\left(\boldsymbol{\theta},\boldsymbol{\theta}^{\left(m\right)}\right) 
&=& \sum_{l=1}^L \hat{\sz}_{1,l} \log\left(\delta_l\right)+\sum_{l=1}^L\sum_{k=1}^L\sum_{t=1}^T\widehat{\sz\sz}_{t,l,k}\log\left(\sq_{l,k}\right) \nonumber\\ 
&+&\sum_{l=1}^L\sum_{t=0}^T\hat{\sz}_{t,l}\left\{-\frac{p}{2}\log\left(2\pi\right)-\frac{1}{2}\log\vert\boldsymbol{\Sigma}_l\vert\right\}\nonumber\\
&+&\sum_{l=1}^L\sum_{t=0}^T\hat{\sz}_{t,l}\left\{-\frac{\sw_{t,l}}{2}\left({\bf y}_t-\boldsymbol{\mu}_{l}\right)'\boldsymbol{\Sigma}_l^{-1}\left({\bf y}_t-\boldsymbol{\mu}_{l}\right)\right\}\nonumber\\
&+&\sum_{l=1}^L\sum_{t=0}^T\hat{\sz}_{t,l}\left\{\frac{\nu_l}{2}\log\left(\frac{\nu_l}{2}\right)-\log\Gamma\left(\frac{\nu_l}{2}\right)\right\}\nonumber\\
&+&\sum_{l=1}^L\sum_{t=0}^T\hat{\sz}_{t,l}\left\{\frac{\nu_l}{2}\left(\log (\sw_{t,l})-\sw_{t,l}\right)+\left(\frac{p}{2}-1\right)\log\left(\sw_{t,l}\right)\right\},\nonumber
\end{eqnarray}
where the conditional expectations
$\hat{\sz}_{t,l}=\mathbb{E}\left(\sz_{t,l}\mid\by_1,\dots,\by_T\right)$ and $\widehat{\sz\sz}_{t,l,k}=\mathbb{E}\left(zz_{t,l}\mid\by_1,\dots,\by_T\right)$, $\forall t=1,2,\dots,T$ and $\forall l,k=1,2,\dots,L$ are computed via the well-know Forward--Filtering Backward--Smoothing (FFBS) recursive algorithm (see Baum \textit{et al.} \cite{baum_etal.1970}). For an introduction to the FFBS algorithm we refer the reader to the book of Fr\"uhwirth-Schnatter \cite{fruhwirth_schnatter.2006}.
\item[M--step:] at iteration $\left(m+1\right)$, the M--step maximizes the function $\mathcal{Q}\left(\boldsymbol{\theta},\boldsymbol{\theta}^{\left(m\right)}\right)$ with respect to $\boldsymbol{\theta}$ to determine the next set of parameters $\boldsymbol{\theta}^{\left(m+1\right)}$. The updated estimates of the hidden parameters, the mean vector $\boldsymbol{\mu}_l$, and the scale matrix $\boldsymbol{\Sigma}_l$ are given by the following expressions:
\begin{eqnarray}
\delta_l^{\left(m+1\right)}&=&\hat{\sz}_{1,l}^{\left(m\right)}\nonumber\\
\sq_{l,k}^{\left(m+1\right)}&=&\frac{\sum_{t=2}^T\widehat{\sz\sz}_{t,l,k}^{\left(m\right)}}{\sum_{k=1}^L\sum_{t=2}^T\widehat{\sz\sz}_{t,l,k}^{\left(m\right)}}\nonumber\\
\bmu_l^{\left(m+1\right)}&=&\frac{\sum_{t=1}^T\hat{\sz}_{t,l}^{\left(m\right)}\hat{\sw}_{t,l}^{\left(m\right)}\by_t}{\sum_{t=1}^T\hat{\sz}_{t,l}^{\left(m\right)}\hat{\sw}_{t,l}^{\left(m\right)}}\nonumber\\
\bSigma_l^{\left(m+1\right)}&=&\frac{\sum_{t=1}^T\hat{\sz}_{t,l}^{(r)}\hat{\sw}_{t,l}^{\left(m\right)}\left(\by_t-\bmu_l^{\left(m+1\right)}\right)\left(\by_t-\bmu_l^{\left(m+1\right)}\right)^\trasp}{\sum_{t=1}^T\hat{\sz}_{t,l}^{\left(m\right)}\hat{\sw}_{t,l}^{\left(m\right)}}\nonumber
\end{eqnarray}
$\forall l=1,2,\dots,L$, where $\sw_{t,l}^{\left(m\right)}$ denotes the current estimate of the conditional expectation of $\sW_t$ given the observation at time $t$, $\by_t$, and $\sz_{t,l}=1$
\begin{equation}
\hat{\sw}_{t,l}^{\left(m\right)}=\frac{\nu_l^{\left(m\right)}+p}{\nu_l^{\left(m\right)}+\left(\by_t-\bmu_{l}^{\left(m\right)}\right){\bSigma_l^{\left(m\right)}}^{-1}\left(\by_t-\bmu_{l}^{\left(m\right)}\right)},\nonumber
\end{equation}
$\forall t=1,2,\dots,T$ and $\forall l=1,2,\dots,L$. The updated estimate $\nu_l^{\left(m+1\right)}$ does not exist in closed form but is given as the solution of the equation
\begin{eqnarray}
&&\frac{\sum_{t=1}^T\hat{\sz}_{t,l}\left(\log\left(\hat{\sw}_{t,l}^{\left(m\right)}\right)-\hat{\sw}_{t,l}^{\left(m\right)}\right)}{\sum_{t=1}^T\hat{\sz}_{t,l}}-\psi\left(\frac{\nu_l^{\left(m\right)}}{2}\right)+\log\left(\frac{\nu_l^{\left(m\right)}}{2}\right)+1+\nonumber\\
&&\qquad\qquad\qquad\qquad\qquad\qquad\qquad+\psi\left(\frac{\nu_l^{\left(m\right)}+p}{2}\right)-\log\left(\frac{\nu_l^{\left(m\right)}+p}{2}\right)=0\nonumber
\end{eqnarray}
where $\psi(\cdot)$ is the Digamma function. The solution can be determined by a bisection algorithm or quasi--Newton methods. As an alternative, we adopt the following approximation due to Shoham \cite{shoham.2002}
\begin{eqnarray}
\nu_l^{\left(m+1\right)}=\frac{2}{h_l+\log\left(h_l\right)-1}+a_0\left[1+{\rm erf}\left(a_1\log\left(\frac{a_2}{h_l+\log\left(h_l\right)-1}\right)\right)\right],\nonumber
\end{eqnarray}
with $a_0=0.0416$, $a_1=0.6594$, $a_2=2.1971$, where ${\rm erf\left(\cdot\right)}$ is the error function and 
\begin{equation}
h_l\equiv-\frac{\sum_{t=1}^T\hat{\sz}_{t,l}^{\left(m\right)}\left[
\psi\left(\frac{p+\nu_l^{\left(m\right)}}{2}\right)
+\log\left(\frac{2}{\nu_l^{\left(m\right)}+\left(\by_t-\boldsymbol{\mu}_l^{\left(m\right)}\right){\bSigma_l^{\left(m\right)}}^{-1}\left(\by_t-\boldsymbol{\mu}_l^{\left(m\right)}\right)^\trasp}\right)-\hat{\sw}_{t,l}^{\left(m\right)}
\right]
}{\sum_{t=1}^T\hat{\sz}_{t,l}^{\left(m\right)}},\nonumber
\end{equation}
$\forall l=1,2,\dots,L$.
\end{description}
%
\section*{Appendix B}
\label{sec:appendix_B}
%
\begin{proof2}{{\rm\textbf{Proposition \ref{prop:univmixstudent_TCE}.}}} Let $\sY$ be a univariate Student--t mixture defined as in Proposition \ref{prop:univmixstudent_TCE}, then the $\tau$--level TCE of $\sY$ is
\begin{eqnarray*}
{\rm TCE}_{\mathsf{Y}}\left(\hat{\sy}^\tau\right)
&=&\xp\left(\mathsf{Y}\mid\mathsf{Y}\leq\hat{\sy}^\tau\right)\nonumber\\
&=&\frac{1}{\mathbb{P}\left(\sY\leq\hat{\sy}^\tau\right)}\int_{-\infty}^{\hat{\sy}^\tau}\mathsf{y}\left[\sum_{l=1}^L\eta_l\mathcal{T}\left(\mathsf{y}\mid\mu_l,\sigma^2_l,\nu_l\right)\right]\,d\mathsf{y}\nonumber\\
&=&\sum_{l=1}^L\frac{\eta_l F_\sY\left(\hat{\sy}^\tau,\mu_l,\sigma^2_l,\nu_l\right)}{\mathbb{P}\left(\sY\leq\hat{\sy}^\tau\right)}{\rm TCE}_{\sY,l}\left(\hat{\sy}^\tau,\mu_l,\sigma^2_l,\nu_l\right)\nonumber\\
&=&\sum_{l=1}^L\pi_l{\rm TCE}_{\sY,l}\left(\hat{\sy}^\tau,\mu_l,\sigma^2_l,\nu_l\right)\nonumber
\end{eqnarray*}
where $\mathbb{P}\left(\sY\leq\hat{\sy}^\tau\right)=\sum_{l=1}^LF_\sY\left(\hat{\sy}^\tau,\mu_l,\sigma^2_l,\nu_l\right)$ with $F_\sY\left(\hat{\sy}^\tau,\mu_l,\sigma^2_l,\nu_l\right)$ and $\pi_l$, $\forall l=1,2,\dots,L$ defined as in Proposition \ref{prop:univmixstudent_TCE}. The TCE of each mixture component ${\rm TCE}_{\sY,l}\left(\hat{\sy}^\tau,\mu_l,\sigma^2_l,\nu_l\right)$, for $l=1,2,\dots,L$ can be evaluated as
\begin{equation}
{\rm TCE}_{\sY,l}\left(\hat{\sy}^\tau,\mu_l,\sigma^2_l,\nu_l\right)=
\frac{1}{F_\sY\left(\hat{\sy}^\tau,\mu_l,\sigma^2_l,\nu_l\right)}
\int_{-\infty}^{\hat{\sy}^\tau}\frac{\sy\Gamma\left(\frac{\nu_l+1}{2}\right)}{\Gamma\left(\frac{\nu_l}{2}\right)\sqrt{\sigma_l^2\pi\nu_l}}\left[
1+\frac{\left(\sy-\mu_l\right)^2}{\nu\sigma_l^2}
\right]^{-\frac{\nu+1}{2}}d\sy\nonumber
\end{equation}
and follows from standard integration results.
\end{proof2}
\begin{proposition}[\textbf{TCE for multivariate Gaussian distributions}]
\label{th:trunc_mean_multiv_norm}
Let $\bY$ be a multivariate Gaussian random variable of dimension $\sd$, i.e. $\bY\sim\mathcal{N}_\sd\left(\bmu,\bLambda\bC\bLambda\right)$ with $\bLambda=\text{Diag}\left\{\sigma_{1},\sigma_{2},\dots,\sigma_{\sd}\right\}$ and correlation matrix $\bC$, then the multivariate tail conditional expectation of $\bY$, i.e. the mean of $\bY$ truncated below the threshold $\widetilde{\by}$, is
\begin{eqnarray}
\label{eq:tce_multivgauss}
{\rm TCE}_{\bY}\left(\widetilde{\by},\bmu,\bLambda, \bC\right)=\bmu+\frac{\bLambda\bC\widehat{\bphi}_{\widetilde{\bz}}}{\Phi\left(\widetilde{\bz}\right)}
\end{eqnarray}
with
\begin{eqnarray}
\label{eq:multiv_trunc_indep_normal_mean}
\widehat{\bphi}_{\widetilde{\bz}}=\left[
\begin{array}{c}
\phi\left(\hat{\mathsf{z}}_{1}\right)\Phi_{-1}\left(\tilde{\mathsf{z}}_{1}\right)\\
\phi\left(\hat{\mathsf{z}}_{2}\right)\Phi_{-2}\left(\tilde{\mathsf{z}}_{2}\right)\\
\vdots\\
\phi\left(\hat{\mathsf{z}}_{\sd}\right)\Phi_{-\sd}\left(\widetilde{\mathsf{z}}_{\sd}\right)\\
\end{array}\right]
\end{eqnarray}
where $\widetilde{\bz}=\bLambda^{-1}\left(\widetilde{\by}-\bmu\right)$,  $\phi\left(\cdot\right)$ denotes the pdf of the standardized Gaussian distribution and 
\begin{equation}
\Phi_{-j}\left(\tilde{\mathsf{z}}_{j}\right)=\int_{\bz\leq\tilde{\bz}} 
\phi\left(\bz\right)\,d\bz,\qquad\forall j=1,2,\dots,\sd.
\end{equation}
and $\bar{\bz}=\bOmega_{22,j}^{-\frac{1}{2}}\left(\bar{\bz}_{2,j}-\bOmega_{22,j}\bC_{22,j}^{-1}\bC_{21,j}\bC_{1|2,j}^{-1}\tilde{\mathsf{z}}_j\right)$, $\bOmega_{22,j}=\left[\bC_{22,j}-\bC_{21,j}\bC_{12,j}\right]^{-1}$.
\end{proposition}
\begin{proof2}
Let $\bZ\sim\mathcal{N}_\sd\left(0,\bC\right)$ be a $\sd$--dimensional Gaussian random variable, consider $\bY=\bmu+\bLambda\bZ$, then $\bY$ has a Gaussian distribution, i.e. $\bY\sim\mathcal{N}_\sd\left(\bmu,\bLambda\bC\bLambda\right)$, then the TCE of $\bY$ is
\begin{eqnarray}
\label{eq:tcerelations}
{\rm TCE}_{\bY}\left(\widetilde{\by},\bmu,\bLambda,\bC\right)
&=&\xp\left(\bY\mid\bY\leq \widetilde{\by}\right)\nonumber\\
&=&\bmu+\bLambda\xp\left(\bZ\mid\bZ\leq\widetilde{\bz}\right)\nonumber\\
&=&\bmu+\bLambda{\rm TCE}_{\bZ}\left(\widetilde{\bz},\bC\right),
\end{eqnarray}
where ${\rm TCE}_{\bZ}\left(\widetilde{\bz},\bC\right)$, is the TCE of the Gaussian distribution $\bZ\sim\mathcal{N}_\sd\left(0,\bC\right)$, and can be evaluated as follows
\begin{eqnarray}
{\rm TCE}_{\bZ}\left(\widetilde{\bz},\bC\right)=\xp\left(\bz,\bz\leq\widetilde{\bz}\right)\equiv
\left[
\begin{array}{c}
\xp\left(\sz_1,\bz\leq\widetilde{\bz}\right)\\
\xp\left(\sz_2,\bz\leq\widetilde{\bz}\right)\\
\vdots\\
\xp\left(\sz_\sd,\bz\leq\widetilde{\bz}\right)
\end{array}
\right].
\end{eqnarray}
Let us consider $\xp\left(\sz_j,\bz\leq\widetilde{\bz}\right)$, $\forall j=1,2,\dots,\sd$, we have
\begin{eqnarray}
\xp\left(\sz_j,\bz\leq\widetilde{\bz}\right)&=&\int_{\sZ_{-j}\leq\tilde{\sz}_{-j}}\underbrace{\left[\int_{\bZ_j\leq\tilde{\bz}_j}
\sz_{j}\phi\left(\sz_j,\mu_{\left[j\vert-j\right]},\sigma^2_{\left[j\vert-j\right]}\right)d\sz_j\right]}_{Integral\,\bA}\nonumber\\
&&\qquad\qquad\qquad\qquad\times\phi_{\sd-1}\left(\bz_{-j},0,{\rm C}_{-j,-j}\right)d\bz_{-j},
\label{eq:tce_z_j_main_expression}
\end{eqnarray}
where $\mu_{\left[j\vert-j\right]}={\rm C}_{\left[j,-j\right]}{\rm C}_{\left[-j,-j\right]}^{-1}\bz_j$, and $\sigma^2_{\left[j\vert-j\right]}=1-{\rm C}_{\left[j,-j\right]}{\rm C}_{\left[-j,-j\right]}^{-1}{\rm C}_{\left[-j,j\right]}$. 
Let us now consider the integral \textbf{A}:
\begin{eqnarray}
\int_{\sZ_j\leq\tilde{\sz}_j}
\sz_{j}\phi\left(\sz_j,\mu_{\left[j\vert-j\right]},\sigma^2_{\left[j\vert-j\right]}\right)d\sz_j,\nonumber
\end{eqnarray}
applying transformation $t=\frac{\sz_j-\mu_{\left[j\vert-j\right]}}{\sigma_{\left[j\vert-j\right]}}$ we get:
\begin{eqnarray}
\int_{\sZ_j\leq\tilde{\sz}_j}
\sz_{j}\phi\left(\sz_j,\mu_{\left[j\vert-j\right]},\sigma^2_{\left[j\vert-j\right]}\right)d\sz_j&=&-\sigma_{\left[j\vert-j\right]}\phi\left(\frac{\tilde{\sz}_j-\mu_{\left[j\vert-j\right]}}{\sigma_{\left[j\vert-j\right]}},0,1\right)\nonumber\\
&&\qquad\quad+{\rm C_{j,-j}{\rm C}_{-j,-j}^{-1}}\bz_{-j}\Phi\left(\frac{\tilde{\sz}_{j}-\mu_{\left[j\vert-j\right]}}{\sigma_{\left[j\vert-j\right]}},0,1\right).
\nonumber
\end{eqnarray}
Plugging this last expression into equation (\ref{eq:tce_z_j_main_expression}) we obtain
\begin{eqnarray}
\xp\left(\sz_j,\bz\leq\widetilde{\bz}\right)&=&-\sigma_{\left[j\vert-j\right]}\underbrace{\int_{\bZ_{-j}\leq\tilde{\bz}_{-j}}
\phi\left(\frac{\tilde{\sz}_j-\mu_{\left[j\vert-j\right]}}{\sigma_{\left[j\vert-j\right]}}\right)\phi_{\sd-1}\left(\bz_{-j},0,{\rm C}_{\left[-j,-j\right]}\right)d\bz_{-j}}_{{\rm Integral}\,\mathbf{B}}\nonumber\\
&&\qquad\quad+{\rm C_{j,-j}{\rm C}_{-j,-j}^{-1}}\int_{\bZ_{-j}\leq\tilde{\bz}_{-j}}\bz_{-j}\Phi\left(\frac{\tilde{\sz}_{j}-\mu_{\left[j\vert-j\right]}}{\sigma_{\left[j\vert-j\right]}},0,1\right)\nonumber\\
&&\qquad\qquad\qquad\qquad\qquad\qquad\qquad\times\phi_{\sd-1}\left(\bz_{-j},0,{\rm C}_{\left[-j,-j\right]}\right)d\bz_{-j}\nonumber.
\label{eq:tce_z_j_second_expression_from_main}
\end{eqnarray}
%
%
%
%
Considering now the integral $\bB$
\begin{eqnarray}
&&\int_{\bZ_{-j}\leq\tilde{\bz}_{-j}}
\phi\left(\frac{\tilde{\sz}_j-\mu_{\left[j\vert-j\right]}}{\sigma_{\left[j\vert-j\right]}}\right)\phi_{\sd-1}\left(\bz_{-j},0,{\rm C}_{\left[-j,-j\right]}\right)d\bz_{-j}\nonumber\\
&&\qquad\qquad=\int_{\bZ_{-j}\leq\tilde{\bz}_{-j}}\frac{1}{\sqrt{2\pi}}\exp\left\{
\frac{\left(\tilde{\sz}_j-\mu_{\left[j\vert-j\right]}\right)^2}{\sigma_{\left[j\vert-j\right]}^2}
\right\}\nonumber\\
&&\qquad\qquad\times\frac{1}{\left(2\pi\right)^{\frac{\sd-1}{2}}}\frac{1}{\vert{\rm C}_{\left[-j,-j\right]}\vert^{\frac{1}{2}}}
\exp\left\{-\frac{1}{2}\bz_{-j}^\trasp{\rm C}_{\left[-j,-j\right]}^{-1}\bz_{-j}
\right\}d\bz_{-j},
\end{eqnarray}
and completing the square in the following way
\begin{eqnarray}
&&\bz_{-j}^\trasp{\rm C}_{\left[-j,-j\right]}^{-1}\bz_{-j}\nonumber\\
&&\qquad+\left(
\tilde{\sz}_j-{\rm C}_{\left[j,-j\right]}{\rm C}_{\left[-j,-j\right]}^{-1}\bz_{-j}
\right)^\trasp\sigma_{\left[j\vert-j\right]}^{-1}
\left(
\tilde{\sz}_j-{\rm C}_{\left[j,-j\right]}{\rm C}_{\left[-j,-j\right]}^{-1}\bz_{-j}
\right)\nonumber\\
&&\qquad=\tilde{\sz}_j^\trasp\tilde{\sz}_j+\left(
\bz_{-j}-{\rm C}_{\left[-j\vert j\right]}{\rm C}_{\left[-j,-j\right]}^{-1}{\rm C}_{\left[-j,j\right]}\sigma_{\left[j\vert-j\right]}^{-1}\tilde{\sz}_j
\right)^\trasp
{\rm C}_{\left[-j\vert j\right]}^{-1}\nonumber\\
&&\qquad\qquad\qquad\qquad\times\left(
\bz_{-j}-{\rm C}_{\left[-j\vert j\right]}{\rm C}_{\left[-j,-j\right]}^{-1}{\rm C}_{\left[-j,j\right]}\sigma_{\left[j\vert-j\right]}^{-1}\tilde{\sz}_j
\right)
\end{eqnarray}
it becomes
{\footnotesize
\begin{eqnarray}
&&\int_{\bZ_{-j}\leq\bz_{-j}}
\phi\left(\frac{\tilde{\sz}_j-\mu_{\left[j\vert-j\right]}}{\sigma_{\left[j\vert-j\right]}}\right)\phi_{\sd-1}\left(\bz_{-j},0,{\rm C}_{\left[-j,-j\right]}\right)d\bz_{-j}\nonumber\\
&&\qquad=
\left(2\pi\right)^{-\frac{\sd}{2}}\vert{\rm C}_{\left[-j,-j\right]}\vert^{-\frac{1}{2}}
\int_{\bZ_{-j}\leq\tilde{\bz}_{-j}}\exp\left\{-\frac{1}{2}\left(
\bz_{-j}-{\rm C}_{\left[-j\vert j\right]}{\rm C}_{\left[-j,-j\right]}^{-1}{\rm C}_{\left[-j,j\right]}{\rm C}_{\left[j\vert -j\right]}^{-1}\tilde{\sz}_j
\right)^\trasp{\rm C}_{\left[-j\vert j\right]}^{-1}\right.\nonumber\\
&&\qquad\qquad\qquad\qquad\qquad\qquad\qquad\qquad\qquad\times\left.\left(
\bz_{-j}-{\rm C}_{\left[-j\vert j\right]}{\rm C}_{\left[-j,-j\right]}^{-1}{\rm C}_{\left[-j,j\right]}{\rm C}_{\left[j\vert -j\right]}^{-1}\tilde{\sz}_j
\right)
\right\}d\bz_{-j}\nonumber\\
&&\qquad=\frac{\phi\left(\tilde{\sz}_j\right)}{\left(2\pi\right)^{\frac{\sd-1}{2}}\vert{\rm C}_{\left[-j,-j\right]}\vert^{\frac{1}{2}}}
\int_{\bZ_{-j}\leq\tilde{\bz}_{-j}}\exp\left\{-\frac{1}{2}\left(
\bz_{-j}-{\rm C}_{\left[-j\vert j\right]}{\rm C}_{\left[-j,-j\right]}^{-1}{\rm C}_{\left[-j,j\right]}{\rm C}_{\left[j\vert -j\right]}^{-1}\tilde{\sz}_j
\right)^\trasp{\rm C}_{\left[-j\vert j\right]}^{-1}\right.\nonumber\\
&&\qquad\qquad\qquad\qquad\qquad\qquad\qquad\qquad\qquad\times\left.\left(
\bz_{-j}-{\rm C}_{\left[-j\vert j\right]}{\rm C}_{\left[-j,-j\right]}^{-1}{\rm C}_{\left[-j,j\right]}{\rm C}_{\left[j\vert -j\right]}^{-1}\tilde{\sz}_j
\right)
\right\}d\bz_{-j}.\nonumber
\end{eqnarray}}
\par\noindent Considering the transformation $t={\rm C}_{\left[-j\vert j\right]}^{-\frac{1}{2}}\left(\bz_{-j}-{\rm C}_{\left[-j\vert j\right]}{\rm C}_{\left[-j,-j\right]}^{-1}{\rm C}_{\left[-j,j\right]}{\rm C}_{\left[-j\vert j\right]}^{-1}\tilde{\sz}_{j}\right)$ and defining $\tilde{t}\equiv{\rm C}_{\left[-j\vert j\right]}^{-\frac{1}{2}}\left(\tilde{\bz}_{-j}-{\rm C}_{\left[-j\vert j\right]}{\rm C}_{\left[-j,-j\right]}^{-1}{\rm C}_{\left[-j,j\right]}{\rm C}_{\left[-j\vert j\right]}^{-1}\tilde{\sz}_{j}\right)$, the previous integral $\bB$ becomes
{\footnotesize
\begin{eqnarray}
&&\int_{\bZ_{-j}\leq\tilde{\bz}_{-j}}
\phi\left(\frac{\tilde{\sz}_j-\mu_{\left[j\vert-j\right]}}{\sigma_{\left[j\vert-j\right]}}\right)\phi_{\sd-1}\left(\bz_{-j},0,{\rm C}_{\left[-j,-j\right]}\right)d\bz_{-j}\nonumber\\
&&\qquad=\frac{\phi\left(\tilde{\sz_j}\right)}{\vert{\rm C}_{\left[-j,-j\right]}\vert^{\frac{1}{2}}}\vert{\rm C}_{\left[-j\vert j\right]}\vert^{\frac{1}{2}}\int_{t\leq\tilde{t}}\frac{\exp\left\{\frac{1}{2}t^\trasp t\right\}}{\left(2\pi\right)^{\frac{\sd-1}{2}}}dt\nonumber\\
&&\qquad=\frac{\phi\left(\tilde{\sz_j}\right)}{\vert{\rm C}_{\left[-j,-j\right]}\vert^{\frac{1}{2}}}\vert{\rm C}_{\left[-j\vert j\right]}\vert^{\frac{1}{2}}\Phi\left({\rm C}_{\left[-j\vert j\right]}^{-\frac{1}{2}}\left(\tilde{\bz}_{-j}-{\rm C}_{\left[-j\vert j\right]}{\rm C}_{\left[-j,-j\right]}^{-1}{\rm C}_{\left[-j,j\right]}{\rm C}_{\left[-j\vert j\right]}^{-1}\tilde{\sz}_{j}\right)\right).
\end{eqnarray}}
Let now consider the last part of integral in equation \eqref{eq:tce_z_j_second_expression_from_main}
{\footnotesize
\begin{eqnarray}
&&\int_{\bZ_{-j}\leq\tilde{\bz}_{-j}}\bz_{-j}\Phi\left(\frac{\tilde{\sz}_{j}-\mu_{\left[j\vert-j\right]}}{\sigma_{\left[j\vert-j\right]}},0,1\right)\phi_{\sd-1}\left(\bz_{-j},0,{\rm C}_{\left[-j,-j\right]}\right)d\bz_{-j}\nonumber\\
&&\qquad=\int_{\bZ_{-j}\leq\tilde{\bz}_{-j}}\bz_{-j}\int_{\sz_{j}\leq\frac{\tilde{\sz}_{j}-\mu_{\left[j\vert-j\right]}}{\sigma_{\left[j\vert-j\right]}}}
\frac{1}{\sqrt{2\pi}}\exp\left\{-\frac{1}{2}t^2\right\}\frac{1}{\left(2\pi\right)^{\frac{\sd-1}{2}}}\exp\left\{-\frac{1}{2}\bz_{-j}^\trasp{\rm C}_{\left[-j,-j\right]}^{-1}\bz_{-j}\right\}d\sz_{j}d\bz_{-j}\nonumber\\
&&\qquad=\int_{\bZ_{-j}\leq\tilde{\bz}_{-j}}\bz_{-j}\int_{\sZ_{j}\leq\tilde{\sz}_j}
\frac{1}{\sigma_{\left[j\vert-j\right]}\sqrt{2\pi}}\exp\left\{-\frac{\left(\tilde{\sz}_{j}-\mu_{\left[j\vert-j\right]}\right)^2}{2\sigma^2_{\left[j\vert-j\right]}}\right\}\nonumber\\
&&\qquad\qquad\qquad\qquad\qquad\qquad\qquad\times\frac{1}{\left(2\pi\right)^{\frac{\sd-1}{2}}}\exp\left\{-\frac{1}{2}\bz_{-j}^\trasp{\rm C}_{\left[-j,-j\right]}^{-1}\bz_{-j}\right\}d\sz_{j}d\bz_{-j}\nonumber\\
&&\qquad=\frac{1}{\sigma_{\left[j\vert-j\right]}}\int_{\bZ\leq\tilde{\bz}}\frac{\bz_{-j}}{\left(2\pi\right)^{\frac{\sd}{2}}}
\exp\left\{-\frac{1}{2}\bz^\trasp{\rm C}^{-1}\bz\right\}d\bz\nonumber\\
&&\qquad=\frac{1}{\sigma_{\left[j\vert-j\right]}}\int_{\bZ\leq\tilde{\bz}}\frac{\bz_{-j}}{\left(2\pi\right)^{\frac{\sd}{2}}}
\exp\left\{-\frac{1}{2}\bz^\trasp{\rm C}^{-1}\bz\right\}d\bz\nonumber\\
&&\qquad=\frac{\vert{\rm C}\vert^{\frac{1}{2}}}{\sigma_{\left[j\vert-j\right]}}\int_{\bZ\leq\tilde{\bz}}\frac{\bz_{-j}\exp\left\{-\frac{1}{2}\bz^\trasp{\rm C}^{-1}\bz\right\}}{\vert{\rm C}\vert^{\frac{1}{2}}\left(2\pi\right)^{\frac{\sd}{2}}}
d\bz\nonumber\\
&&\qquad=\frac{\vert{\rm C}\vert^{\frac{1}{2}}}{\sigma_{\left[j\vert-j\right]}}\xp\left(\bz_{-j},\bZ\leq\tilde{\bz}\right)\nonumber\\
&&\qquad=\vert{\rm C}_{\left[-j,-j\right]}\vert^{\frac{1}{2}}\xp\left(\bz_{-j},\bZ\leq\tilde{\bz}\right).\nonumber
\end{eqnarray}}
because $\vert{\rm C}\vert=\vert{\rm C}_{\left[-j,-j\right]}\vert\sigma^2_{\left[j\vert-j\right]}$.
Concluding, we have that:
{\footnotesize
\begin{eqnarray}
&&\xp\left(\sz_j,\bz\leq\tilde{\bz}\right)=-\sigma_{\left[j,-j\right]}\phi\left(\tilde{\sz}_j\right)\frac{\vert{\rm C}_{\left[-j\vert j\right]}\vert^{\frac{1}{2}}}{\vert{\rm C}_{\left[-j\vert-j\right]}\vert^{\frac{1}{2}}}\Phi\left({\rm C}_{\left[-j\vert j\right]}^{-\frac{1}{2}}\left(\tilde{\bz}_{-j}-{\rm C}_{\left[-j\vert j\right]}{\rm C}_{\left[-j,-j\right]}^{-1}{\rm C}_{\left[-j,j\right]}{\rm C}_{\left[-j\vert j\right]}^{-1}\tilde{\sz}_{j}\right)\right)\nonumber\\
&&\qquad\qquad\qquad\qquad\quad+{\rm C}_{\left[j,-j\right]}{\rm C}_{\left[-j,-j\right]}^{-1}\vert{\rm C}_{\left[-j,-j\right]}\vert^{\frac{1}{2}}\xp\left(\bz_{-j},\bZ\leq\tilde{\bz}\right),
\label{eq:TCE_final_before_system}
\end{eqnarray}}
for $j=1,2,\ldots,\sd$. Let 
\begin{eqnarray}
\hat{\sz}_j&\equiv&\xp\left(\sz_j,\bz\leq\tilde{\bz}\right)\nonumber\\
\hat{\sz}_{-j}&\equiv&\xp\left(\sz_{-j},\bz\leq\tilde{\bz}\right)\nonumber\\
\end{eqnarray}
rewriting the previous equation (\ref{eq:TCE_final_before_system}) as
{\footnotesize
\begin{eqnarray}
&&\hat{\sz}_j-{\rm C}_{\left[j,-j\right]}{\rm C}_{\left[-j,-j\right]}^{-1}\vert{\rm C}_{\left[-j,-j\right]}\vert^{\frac{1}{2}}\hat{\sz}_{-j}=\nonumber\\
&&\qquad\qquad\qquad-\sigma_{\left[j,-j\right]}\phi\left(\tilde{\sz}_j\right)\frac{\vert{\rm C}_{\left[-j\vert j\right]}\vert^{\frac{1}{2}}}{\vert{\rm C}_{\left[-j\vert-j\right]}\vert^{\frac{1}{2}}}\Phi\left({\rm C}_{\left[-j\vert j\right]}^{-\frac{1}{2}}\left(\tilde{\bz}_{-j}-{\rm C}_{\left[-j\vert j\right]}{\rm C}_{\left[-j,-j\right]}^{-1}{\rm C}_{\left[-j,j\right]}{\rm C}_{\left[-j\vert j\right]}^{-1}\tilde{\sz}_{j}\right)\right)\nonumber\\
\label{eq:TCE_final_before_system}
\end{eqnarray}}
\par\noindent we get a system of $\sd$ equation with $\sd$ unknowns $\bA\hat{\bz}=\bb$ where the matrix $\bA$ has diagonal elements ${\rm a}_{i,i}=1$, for $i=1,2,\dots,\sd$, and off--diagonal elements ${\rm a}_{i,j}$ and $i\neq j$ being the $j$--th element of the  $\left(\sd-1\right)$--dimensional vector $-{\rm C}_{\left[i,-i\right]}{\rm C}_{\left[-i,-i\right]}^{-1}\vert{\rm C}_{\left[i,-i\right]}\vert^{\frac{1}{2}}$, and the vector $\bb$ has the $j$--th generic element equal to
%
%
%
\begin{equation}
b_j=-\sigma_{\left[j,-j\right]}\phi\left(\tilde{\sz}_j\right)\frac{\vert{\rm C}_{\left[-j\vert j\right]}\vert^{\frac{1}{2}}}{\vert{\rm C}_{\left[-j\vert-j\right]}\vert^{\frac{1}{2}}}\Phi\left({\rm C}_{\left[-j\vert j\right]}^{-\frac{1}{2}}\left(\tilde{\bz}_{-j}-{\rm C}_{\left[-j\vert j\right]}{\rm C}_{\left[-j,-j\right]}^{-1}{\rm C}_{\left[-j,j\right]}{\rm C}_{\left[-j\vert j\right]}^{-1}\tilde{\sz}_{j}\right)\right)
\end{equation}
for $j=1,2,\dots,\sd$. Solving the previous system of equations completes the proof. Without loss of generality we can consider the case where $\sd=2$ where
\begin{eqnarray}
\hat{\bz}=\xp\left(\bz\mid\bz\leq\tilde{\bz}\right)&=&\bA^{-1}\bb\nonumber\\
&=&\frac{1}{1-\rho^2}\left[
\begin{array}{cc}
1 & \rho\\
\rho & 1
\end{array}
\right]
\left[
\begin{array}{c}
-\left(1-\rho^2\right)\phi\left(\tilde{\sz}_1\right)\Phi\left(\frac{\tilde{\sz}_2-\rho\tilde{\sz}_1}{\sqrt{1-\rho^2}}\right)\\
-\left(1-\rho^2\right)\phi\left(\tilde{\sz}_2\right)\Phi\left(\frac{\tilde{\sz}_1-\rho\tilde{\sz}_2}{\sqrt{1-\rho^2}}\right)
\end{array}
\right]\nonumber\\
&=&\left[
\begin{array}{c}
-\phi\left(\tilde{\sz}_1\right)\Phi\left(\frac{\tilde{\sz}_2-\rho\tilde{\sz}_1}{\sqrt{1-\rho^2}}\right)-\rho\phi\left(\tilde{\sz}_2\right)\Phi\left(\frac{\tilde{\sz}_1-\rho\tilde{\sz}_2}{\sqrt{1-\rho^2}}\right)\\
-\phi\left(\tilde{\sz}_2\right)\Phi\left(\frac{\tilde{\sz}_1-\rho\tilde{\sz}_2}{\sqrt{1-\rho^2}}\right)-\rho\phi\left(\tilde{\sz}_1\right)\Phi\left(\frac{\tilde{\sz}_2-\rho\tilde{\sz}_1}{\sqrt{1-\rho^2}}\right)
\end{array}
\right]\nonumber\\
&=&\bC\left[\begin{array}{c}
-\phi\left(\tilde{\sz}_1\right)\Phi\left(\frac{\tilde{\sz}_2-\rho\tilde{\sz}_1}{\sqrt{1-\rho^2}}\right)\\
-\phi\left(\tilde{\sz}_2\right)\Phi\left(\frac{\tilde{\sz}_1-\rho\tilde{\sz}_2}{\sqrt{1-\rho^2}}\right)
\end{array}\right],
\end{eqnarray}
with $\rho={\rm C}_{\left[1,2\right]}{\rm C}_{\left[2,2\right]}^{-1}\vert{\rm C}_{\left[1,2\right]}\vert^{\frac{1}{2}}$.
\end{proof2}
\begin{proposition}[\textbf{TCE for multivariate Student--t distributions}]
\label{th:trunc_mean_multiv_student}
Let $\bY$ be a multivariate Student-t random variable, i.e. $\bY\sim\mathcal{T}_d\left(\bmu,\bLambda\bC\bLambda,\nu\right)$ with $\bLambda=\text{Diag}\left\{\sigma_{1},\sigma_{2},\dots,\sigma_{d}\right\}$, correlation matrix $\bC$ and degrees of freedom $\nu$, then the multivariate tail conditional expectation of $\bY$ is
\begin{eqnarray}
\label{eq:tce_asn}
\mathbb{E}\left(\bY,\bY\leq\widetilde{\by}\right)=\bmu+\bLambda\bC\widehat{\bphi}_{\widetilde{\bz}}
\end{eqnarray}
with
\begin{eqnarray}
\label{eq:multiv_trunc_indep_normal_mean}
\widehat{\bphi}_{\widetilde{\bz}}=\left[
\begin{array}{c}
\phi\left(\hat{\mathsf{z}}_{1}\right)\Phi_{-1}\left(\tilde{\mathsf{z}}_{1}\right)\\
\phi\left(\hat{\mathsf{z}}_{2}\right)\Phi_{-2}\left(\tilde{\mathsf{z}}_{2}\right)\\
\vdots\\
\phi\left(\hat{\mathsf{z}}_{d}\right)\Phi_{-d}\left(\tilde{\mathsf{z}}_{d}\right)\\
\end{array}\right]
\end{eqnarray}
where $\widetilde{\bz}=\bLambda^{-1}\left(\widetilde{\by}-\bmu\right)$,  $\phi\left(\right)$ denotes the pdf of the standardized Student--t distribution and 
\begin{equation}
\Phi_{-j}\left(\tilde{\mathsf{z}}_{j}\right)=\int_{\bz\leq\tilde{\bz}} 
\phi\left(\bz\right)\,d\bz,\qquad\forall j=1,2,\dots ,d.
\end{equation}
and $\bar{\bz}=\bOmega_{22,j}^{-\frac{1}{2}}\left(\bar{\bz}_{2,j}-\bOmega_{22,j}\bC_{22,j}^{-1}\bC_{21,j}\bC_{1|2,j}^{-1}\tilde{\mathsf{z}}_j\right)$, $\bOmega_{22,j}=\left[\bC_{22,j}-\bC_{21,j}\bC_{12,j}\right]^{-1}$.
\end{proposition}
\begin{proof2}
The proof is exactly as that reported for the Gaussian case with the only execption that here we exploit the scale representation of the Sutent--t distribution in equation \eqref{eq:student_scale_mixture}.
\end{proof2}

\begin{proposition}[\textbf{TCE for multivariate Gaussian and Student--t mixtures}] 
Let $\bY$ be a multivariate Gaussian (or Student--t) mixture, i.e. $\bY\sim\sum_{l=1}^L\eta_l \mathcal{N}\left(\by\mid\bmu_l,\bSigma_l,\right)$ $\left(or\,\bY\sim\sum_{l=1}^L\eta_l \mathcal{T}\left(\by\mid\bmu_l,\bSigma_l,\nu_l\right)\right)$, then the tail conditional expectation of $\bY$ is a convex linear combination of the tail conditional expectations of the components:
\begin{eqnarray}
{\rm TCE}_{\bY}\left(\widetilde{\by}, L\right)=\sum_{l=1}^L\pi_l{\rm TCE}_l\left(\widetilde{\by}\right)
\label{eq:univmixasn_tce}
\end{eqnarray}
where the weights are $\pi_l=\eta_l\frac{\Phi\left(\widetilde{\by},\bmu_l,\bSigma_l\right)}{\sum_{l=1}^L\eta_l\Phi\left(\widetilde{\by},\bmu_l,\bSigma_l\right)}$ in the Gaussian case, and $\pi_l=\eta_l\frac{t\left(\widetilde{\by},\bmu_l,\bSigma_l,\nu_l\right)}{\sum_{l=1}^L\eta_lt\left(\widetilde{\by},\bmu_l,\bSigma_l,\nu_l\right)}$ in the Student--t case, $l=1,2,\dots,L$, with $\sum_{l=1}^L\pi_l=1$ and $\Phi\left(\cdot\right)$ and $t\left(\cdot\right)$ denotes the Gaussian and Student--t cdf, respectively.%
\end{proposition}

\begin{proof2}
See Bernardi \cite{bernardi.2013} for a similar proof involving Skew Normal mixtures.
\end{proof2}
%
%
\par\noindent\textbf{Acknowledgements.} This research is supported by the Italian Ministry of Research PRIN 2013--2015, ``Multivariate Statistical Methods for Risk Assessment'' (MISURA), by the ``Carlo Giannini Research Fellowship'', the ``Centro Interuniversitario di Econometria'' (CIdE) and ``UniCredit Foundation'', and by the 2011 Sapienza University of Rome Research Project.
%
\newpage



\begin{thebibliography}{plain}
%
\bibitem{acerbi.2002} \textsc{Acerbi, C., (2002).} \newblock Spectral measures of risk: A coherent representation of
subjective risk aversion, \newblock{\em Journal of Banking \& Finance}, 26, pp. 1505--1518.
%
\bibitem{acerbi_tasche.2002} \textsc{Acerbi, C. and Tasche, D., (2002).} \newblock On the coherence of expected shortfall, \newblock{\em Journal of Banking \& Finance}, 26, pp. 1487--1503
%
\bibitem{acharya_richardson.2009} \textsc{Acharya, V.V., and Richardson, M., (2009).} Restoring financial stability: how to repair a failed system. \newblock John Wiley \& Sons, New York.
%
\bibitem{acharya_etal.2012} \textsc{Acharya, V.V., Engle, R.F. and Richardson, M., (2012).} Capital shortfall: a new approach to ranking and regulating systemic risks. \textit{American Economic Review}, 102, pp. 59--64.
%
\bibitem{acharya_etal.2010} \textsc{Acharya, V.V., E., Philippon, T. and Richardson, M., (2010).} Measuring systemic Risk. \textit{New York University Working Paper}.
%
\bibitem{adrian_brunnermeier.2011} \textsc{Adrian, T. and Brunnermeier, M.K., (2011).} \newblock CoVaR, \newblock{\em Working Paper}, Federal Reserve Bank of New York. 
%
\bibitem{amisano_geweke.2011} \textsc{Amisano, G. and Geweke, J., (2004).} \newblock Hierarchical Markov Normal Mixture models with applications to financial asset returns, \newblock{\em Journal of Applied Econometrics}, 26, pp. 1--29.
%
\bibitem{ang_bekaert.2004} \textsc{Ang, A. and Bekaert, G., (2004).} \newblock How Regimes Affect Asset Allocation, \newblock{\em Financial Analysts Journal}, 60, pp. 86--99.
%
\bibitem{artzner_etal.1999} \textsc{Artzner, P., Delbaen, F., Eber, J.-M. and Heath, D., (1999).} \newblock Coherent measures of risk, \newblock{\em Mathematical Finance}, 9, pp. 203--228.
%
\bibitem{bartolucci_farcomeni.2009} \textsc{Bartolucci, F. and Farcomeni, A., (2009).} \newblock A multivariate extension of the dynamic logit model for longitudinal data based on a latent Markov heterogeneity structure, \newblock{\em Journal of the American Statistical Association}, 104, pp. 816--831. 
%
\bibitem{bartolucci_farcomeni.2010} \textsc{Bartolucci, F. and Farcomeni, A., (2010).} \newblock A Note on the Mixture Transition Distribution and Hidden Markov Models, \newblock{\em Journal of Time Series Analysis}, 31, pp. 132--138. 
%
\bibitem{baum_etal.1970} \textsc{Baum, L.E., Petrie, T., Soules, G. and Weiss, N., (1970).} \newblock A maximisation technique occurring in the statistical analysis of probabilistic functions of finite state Markov chains, \newblock{\em The Annals of Mathematical Statistics}, 37, pp. 1554--1563. 
%
\bibitem{bernardi.2013} \textsc{Bernardi, M., (2013).} \newblock Risk Measures for Skew Normal mixtures, \newblock{\em Statistics \& Probability Letters}, 83, pp. 1819--1824.
%
\bibitem{bernardi_etal.2013} \textsc{Bernardi, M., Gayraud, G. and Petrella, L., (2013).} \newblock Bayesian inference for CoVaR, \newblock{{\tt Preprint arXiv:1306.2834v3 [stat.ME]}, \url{http://arxiv.org/abs/1306.2834}}.
%
\bibitem{bialkowski.2003} \textsc{Bialkowski, J., (2003).} Modelling returns on stock indices for Western and Central European stock exchanges--Markov switching approach, {\em Southeastern European Journal of Economics}, 2, pp. 81--100.
%
\bibitem{billio_etal.2012} \textsc{Billio, M., Getmansky, M., Lo, A.W. and Pellizon, L., (2012).} Econometric measures of connectedness and systemic risk in the finance and insurance sectors. \textit{Journal of Financial Economics}, 104, pp. 535--559.
%
%
\bibitem{brownlees_engle.2012} \textsc{Brownlees, C.T. and Engle, R. (2012).} Volatility, correlation and tails for systemic risk measurement. \textit{Working paper}.
%
\bibitem{bulla.2011} \textsc{Bulla, J., (2011).} \newblock Hidden Markov Models with t Components. Increased Persistence and Other Aspects, \newblock{\em Quantitative Finance}, 11, pp. 459--475.
%
\bibitem{bulla_etal.2011} \textsc{Bulla, J., Mergner, S., Bulla. I., Sesbo\"u\'e, A., and Chesneau, C., (2011).} \newblock Markov--switching asset allocation: Do profitable strategies exist?, \newblock{\em Journal of Asset Management}, 12, pp. 310--321.
%
\bibitem{bulla_and_bulla.2006} \textsc{Bulla, J. and Bulla. I., (2006).} \newblock Stylized facts of financial time series and hidden semi-Markov models, \newblock{\em Computational Statistics \& Data Analysis}, 51, pp. 2192--2209.
%
\bibitem{cao.2013} \textsc{Cao, Z., (2013).} \newblock Multi--CoVaR and Shapley value: a systemic risk measure, \newblock{\em Banque de France Working paper}.
%
\bibitem{cappe_etal.2005} \textsc{Capp\'e, O., Moulines, E. and Ryd\'en, T., (2005).}\newblock{\em Inference in Hidden Markov Models}, Springer Series in Statistics, Springer--Verlag, Berlin.
%
\bibitem{castro_ferrari.2013} \textsc{Castro, C. and Ferrari, S., (2014)}, \newblock Measuring and testing for the systemically important financial institutions, \newblock{\em forthcoming Journal of Empirical Finance}.
%
\bibitem{cizek_etal.2011} \textsc{\v{C}i\v{z}ek, P., H\"ardle, W.K. and Weron, R., (2011)},\newblock{\em Statistical Tools for Finance and Insurance}, Second Edition, Springer--Verlag, New York.
%
%
%
\bibitem{demarta_mcneal.2005} \textsc{Demarta, S. and McNeil, A.J., (2005).} \newblock The $t$ copula and related copula, \newblock{\em International Statistical Review}, 73, pp. 111--129.
%
\bibitem{dempster_etal.1977} \textsc{Dempster, A.P., Laird, N.M., Rubin, D.B., (1977).} \newblock Maximum likelihood from incomplete data using the EM algorithm (with discussion), \newblock{\em Journal of the Royal Statistical Society, (series B)}, 39, pp. 1--39.
%
\bibitem{dymarski.2011} \textsc{Dymarski, P., (2011).} \newblock{\em Hidden Markov Models, Theory and Applications}, \newblock Rijeka, HR, Intech, 207--222.
%
%
\bibitem{embrechts_etal.2000} \textsc{Embrechts, P., McNeil, A.J. and Straumann, D., (1999).} \newblock Correlation and dependence in risk management: properties and pitfalls.\newblock In {\rm Risk management: value at risk and beyond}, pp. 176--223, edited by Dempster, M. and Moffatt, H.K., Cambridge University Press, Cambridge.
%
\bibitem{fruhwirth_schnatter.2006} \textsc{Fr\"{u}hwirth-Schnatter, S., (2006).} \newblock{\em Finite Mixture and Markov Switching Models}. Springer Series in Statistics. Springer, New
York.
%
\bibitem{gettinby_etal.2004} \textsc{Gettinby, G.D., Sinclair, C.D., Power, D.M. and Brown, R.A., (2004)}. \newblock An analysis of the distribution of extreme share returns in the uk from 1975 to 2000,  \newblock{\em Journal of Business Finance \& Accounting}, 31, pp. 607--646.
%
\bibitem{geweke_amisano.2010} \textsc{Geweke, J. and Amisano, G., (2010).} \newblock Comparing and evaluating Bayesian predictive distributions of asset returns, \newblock{\em International Journal of Forecasting}, 26, pp. 216--230. 
%
\bibitem{girardi_ergun.2013}\textsc{Girardi, G., Erg\"{u}n, A.T., (2013).} \emph{Systemic risk measurement: Multivariate GARCH estimation of CoVaR}. \textit{Journal of Banking \& Finance}, \url{http://dx.doi.org/10.1016/j.jbankfin.2013.02.027}.
%
%
%
%
\bibitem{hamilton.1989} \textsc{Hamilton, J.D., (1989).} \newblock A New Approach to the Economic Analysis of Nonstationary Time Series and the Business Cycle, \newblock{\em Econometrica}, 57, pp. 357--384.
%
\bibitem{hamilton.1990} \textsc{Hamilton, J.D., (1990).} \newblock Analysis of time series subject to changes in regime, \newblock{\em Journal of Econometrics}, 45, pp. 39--70.
%
\bibitem{harris_kucukukozmen.2001} \textsc{Harris, R.D. and K\"{u}\c{c}\"{u}k\"{o}zmen, C.C., (2001).} \newblock The empirical distribution of uk and us stock returns, \newblock{\em Journal of Business Finance \& Accounting}, 28, pp. 715--740.
%
%
\bibitem{huang_etal.2012} \textsc{Huang, X., Zhou, H. and Zhu, H., (2012).} \newblock Systemic risk contributions, \newblock{\em Journal of Financial Services Research}, 42, pp. 55--83.
%
%
\bibitem{kotz_nadarajah.2004} \textsc{Kotz, S. and Nadarajah, S., (2004).} \newblock{\em Multivariate t Distributions and their Applications}, Cambridge University Press, USA.
%
\bibitem{koylouglu_stoker.2002} \textsc{Koylouglu, U. and Stoker, J., (2002).} \newblock Honour your contribution, \newblock{\em RISK}, 15, pp. 90--94.
%
\bibitem{lagona_picone.2013} \textsc{Lagona, F. and Picone, M., (2013).} \newblock Maximum likelihood estimation of bivariate circular hidden Markov models from incomplete data, \newblock{\em Journal of Statistical Computation and Simulation}, to appear.
%
%
\bibitem{mclachlan_krishnan.2008} \textsc{McLachlan, G. and Krishnan, T., (2008).} \newblock{\em The EM algorithm and extensions}, \newblock Wiley Series in Probability and Statistics, John Wiley \& Sons, New York.
%
\bibitem{mclachlan_peel.2000} \textsc{McLachlan G. and Peel D., (2000).} \newblock{\em Finite Mixture Models}, \newblock Wiley Series in Probability and Statistics, John Wiley \& Sons, New York.
%
\bibitem{mcneil_etal.2005} \textsc{McNeil, A.J., Frey, R. and Embrechts, P., (2005).} \newblock{\em Quantitative Risk Management: Concepts, Techniques and Tools}, \newblock Princeton Series in Finance, Princeton University Press.
%
\bibitem{peel_mclachlan.2000} \textsc{Peel, D. and Mclachlan, C.J., (2000).} \newblock Robust Mixture Modelling using the t-distribution, \newblock{\em Statistics \& Computing}, 10, pp. 335--344.
%
%
\bibitem{ryden.2008} \textsc{Ryd\'en, T., (2008).} \newblock EM versus Markov chain Monte Carlo for estimation of Hidden Markov models: a computational perspective, \newblock{\em Bayesian Analysis}, 3, pp. 659--688.
%
\bibitem{ryden_etal.1998} \textsc{Ryd\'en, T., Ter\"asvirta, T. and Asbrink, S., (1998).} \newblock Stylized facts of daily return series and the hidden markov model, \newblock{\em Journal of Applied Econometrics}, 13, pp. 217--244.
%
%
\bibitem{shoham.2002} \textsc{Shoham, S., (2002).} \newblock Robust Clustering by Deterministic Agglomeration EM of Mixtures of Multivariate t-Distributions, \newblock{\em Pattern Recognition}, 35, pp. 1127--1142.
%
\bibitem{shapley.1953} \textsc{Shapley, L., (1953).} \newblock A value for n--person Games, \newblock{\em Annals of Mathematical Studies}, 28, pp. 307--317.
%
%
\bibitem{sutradhar.1984} \textsc{Sutradhar, B.C., (1984).} Contributions to Multivariate Analysis based on Elliptic T
Model. \newblock {\em Unpublished Ph.D. Thesis. The University of Western Ontario, Canada.}
%
\bibitem{tarashev_etal.2010} \textsc{Tarashev, N., Borio, C. and Tsatsaronis, K., (2010).} \newblock Attributing systemic risk to individual institutions: methodology and policy applications, \newblock{\em BIS working parper} No. 308.
%
\bibitem{zucchini_macdonald.2009} \textsc{Zucchini, W. and MacDonald, I.,
(2009)}. \newblock{\em Hidden Markov Models for Time Series: an Introduction using R}, \newblock Chapman \& Hall/CRC.
%
\end{thebibliography}
\end{document}